\documentclass[aoas]{imsart}

\usepackage{imsart}
\usepackage{mathrsfs}
\usepackage{natbib,graphicx,lscape,longtable}
\usepackage{mathrsfs,amsmath,amsthm,amssymb}
\usepackage{color,epsfig,graphicx,pdfpages}
\usepackage{rotating}
\usepackage{float}
\usepackage{bm}
\usepackage{ulem}
\usepackage{CJK}
\usepackage{natbib}
\usepackage{algorithm}
\usepackage{algorithmic}
\usepackage{booktabs}
\usepackage{bm}
\usepackage{subfigure}
\usepackage{makecell}
\usepackage{array}
\usepackage{mathtools}

\theoremstyle{plain}

\def\beq{\begin{equation}}
	\def\eeq{\end{equation}}
\def\beqr{\begin{eqnarray}}
	\def\eeqr{\end{eqnarray}}
\def\beqrs{\begin{eqnarray*}}
	\def\eeqrs{\end{eqnarray*}}
\def\bet{\begin{theorem}}
	\def\eet{\end{theorem}}
\def\bel{\begin{lemma}}
	\def\eel{\end{lemma}}
\def\bep{\begin{proposition}}
	\def\eep{\end{proposition}}
\def\bg{\begin{figure}[tbph]\begin{center}}
		\def\eg{\end{center}\end{figure}}

\def\bc{\begin{center}}
	\def\ec{\end{center}}

\def\mR{\mathbb{R}}
\def\mS{\mathcal S}

\def\cov{\mbox{cov}}

\def\argmax{\mbox{argmax}}

\usepackage{lastpage}
\usepackage{hyperref}
\startlocaldefs
\theoremstyle{plain}

\newtheorem{theorem}{\textbf{Theorem}}
\newtheorem{proposition}[theorem]{\textbf{Proposition}}
\newtheorem{corollary}[theorem]{\textbf{Corollary}}
\newtheorem{remark}{\textbf{Remark}}
\newtheorem{lemma}[theorem]{Lemma}
\theoremstyle{remark}

\endlocaldefs

\begin{document}

\begin{frontmatter}
\title{Pseudo-Likelihood Ratio Screening based on Network Data with Applications}

\runtitle{Pseudo-Likelihood Ratio Feature Screening for Network Data}

\begin{aug}
\author[A]{\fnms{Wei}~\snm{Hu}\ead[label=e1]{huw@clas.ac.cn}},
\author[B]{\fnms{Danyang}~\snm{Huang}\ead[label=e2]{dyhuang@ruc.edu.cn}}
\and
\author[B]{\fnms{Bo}~\snm{Zhang}\ead[label=e3]{mabzhang@ruc.edu.cn}}
\address[A]{National Science Library(Chengdu), Chinese Academy of Sciences\printead[presep={,\ }]{e1}}

\address[B]{School of Statistics, Renmin University of China\printead[presep={,\ }]{e2,e3}}
\end{aug}

\begin{abstract}
Social network platforms today generate vast amounts of data, including network structures and a large number of user-defined tags, which reflect users' interests. The dimensionality of these personalized tags can be ultra-high, posing challenges for model analysis in targeted preference analysis. Traditional categorical feature screening methods overlook the network structure, which can lead to incorrect feature set and suboptimal prediction accuracy. This study focuses on feature screening for network-involved preference analysis based on ultra-high-dimensional categorical tags. We introduce the concepts of {\it self-related} features and {\it network-related} features, defined as those directly related to the response and those related to the network structure, respectively. We then propose a pseudo-likelihood ratio feature screening procedure that identifies both types of features. Theoretical properties of this procedure under different scenarios are thoroughly investigated. Extensive simulations and real data analysis on Sina Weibo validate our findings.
\end{abstract}


\end{frontmatter}

\section{Introduction}
\subsection{Preference Analysis Problem and Challenge}
\begin{figure}[h]
	\centering
	\includegraphics[width=9cm]{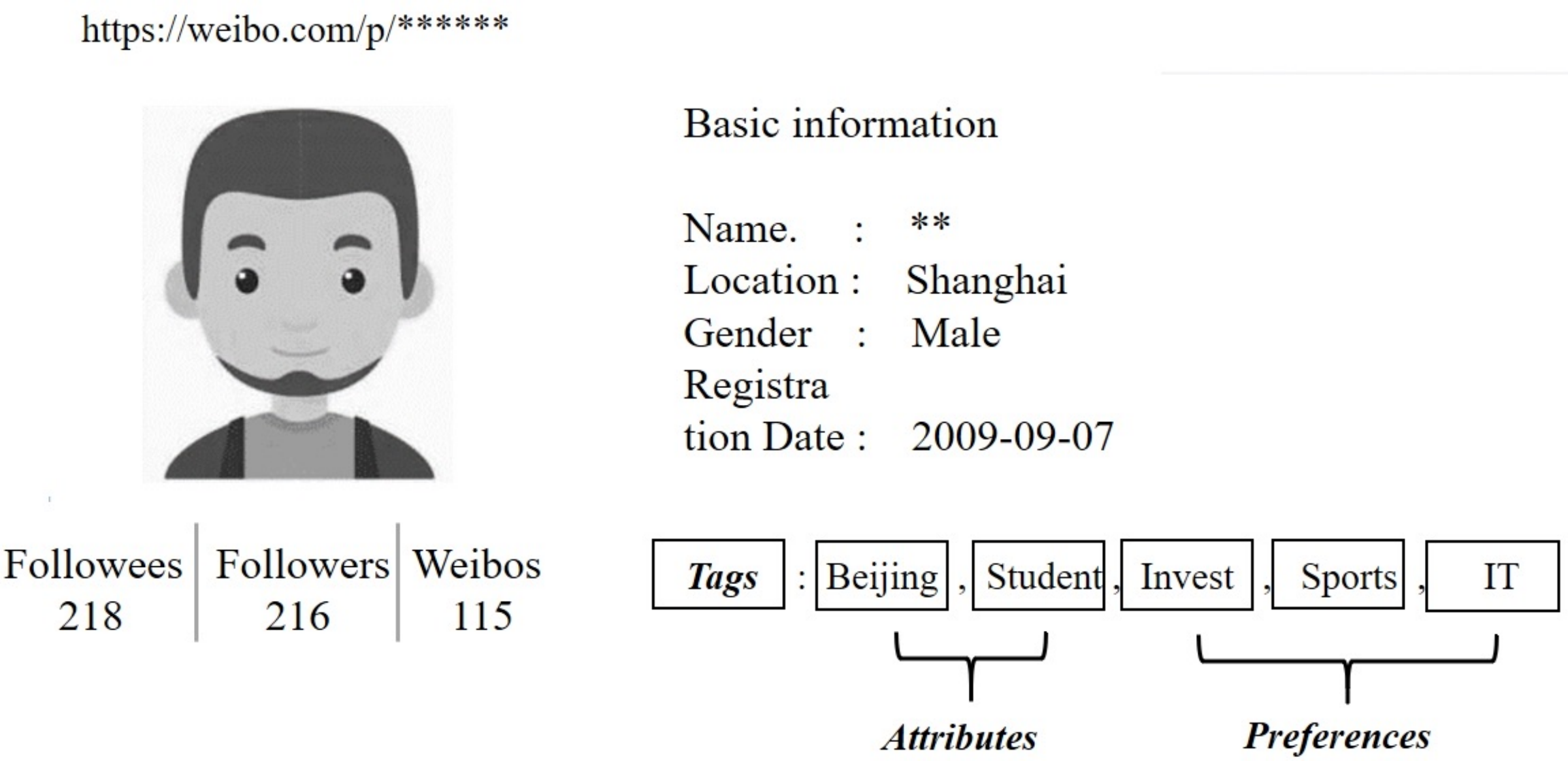}
	\caption{The personal homepage of an anonymous user.}
	\label{weibo}
\end{figure}

In recent years, network analysis has gained significant research interest \citep{wasserman1994social, bruggeman2013social, jin2021survey}. 
 Users' targeted preference analysis based on  network data can drive profit growth through personalized marketing \citep{hagiu2020data, gao2021cross}, customer stickiness \citep{zhang2017influence, maqableh2021effect}, and customer segmentation \citep{foster2011exploring, campbell2014segmenting}. 
A typical example is Sina Weibo, China's largest Twitter-type social network platform, where vast user-defined tags reflect individual identities, habits, and interests (see Figure \ref{weibo}). In this social network, individuals are connected through relationships (edges), forming the network structure. 
An individual's targeted preference (i.e., the response variable) is influenced by (1) their own tags (features) and (2) the tags of their connected users. The application of preference analysis brings about the challenge of feature screening methods designed for classification tasks with network structure involved. This is the focus of this work.

In such data scenarios, the tag dimension can far exceed the number of nodes, resulting in an ultra-high-dimensional classification task. 
Redundant features increase model complexity and introduce additional noise without improving prediction accuracy \citep{li2024feature}.
This issue may be exacerbated by network dependencies, which are overlooked in well-established feature screening methods \citep{mai2013kolmogorov,xie2020category} developed under traditional independence assumptions.
In real-world social networks, studies have shown that users' preferences are often influenced by those of their connections \citep{katona2011network, wang2013modeling, frikha2016time, gao2021cross,milli2025engagement}. 
Existing feature screening methods fail to account for such network structures, which may result in suboptimal prediction performance.
This motivates our proposed feature screening method to address this unique challenge. Finally, from an application perspective, resolving this challenge offers two significant benefits: from the platform’s side, it enables precise customer segmentation and personalized recommendations for advertising or User-Generated Content (UGC); from the user's side, when sharing content based on their preferences, they can select recommended tags to better position their identity, strengthen bonds with followers, and expand their network reach.

\subsection{Related Works and Limitation}

To model the probability of node-level response, network-based autoregressive models have been proposed for continuous response data scenarios \citep{zhu2017network, armillotta2023nonlinear, wu2023random, jiang2023autoregressive, chang2024embedding}. Additionally, \cite{su2019learning} and \cite{sit2021event} explored distance-based dependence structures to describe relationships among individuals. 
Specifically, in the context of the preference analysis problem in network data, researchers focus on classification problems involving network structure \citep{huang2018network, cai2018network, mukherjee2021high}. \cite{huang2018network} incorporated the network structure into the classical na{\"i}ve Bayes model, while \cite{cai2018network} integrated it into linear discriminant analysis. Moreover, \cite{zhang2020logistic} considered network logistic regression based on the similarity of nodal features. \cite{mukherjee2021high} discussed network regression models in high-dimensional settings. \cite{zhang2021depth} further generalized the network logistic regression model by studying data depth, which describes the general association between features and response in network generation probability. 
Compared to traditional classification methods, incorporating network structure significantly improves the accuracy of node-level response predictions due to the additional information derived from individuals’ neighbors' tags. Similar conclusions can also be drawn from graph neural networks \citep{kipf2016semi, ma2019flexible}, linear regression models \citep{li2019prediction, liu2022prediction}, graphical models \citep{li2020high}, and collaborative filtering methods \citep{yu2021collaborative, gao2021cross}.

However, as highlighted earlier, in real-world applications, researchers often collect a large number of categorical tags whose dimensionality far exceeds the number of nodes in the network.
This leads to an ultra-high-dimensional classification problem \citep{fan2009ultrahigh, liu2015selective}. Not all features, however, are informative for predicting the response, even in some deep learning models such as deep forest \citep{pang2020improving}. Consequently, feature screening methods are commonly employed to eliminate redundant features prior to model fitting, a topic that has been well-studied in independent settings since the pioneering work of \citep{fan2008sure}. Various generalized screening methods have been developed in different contexts \citep{li2012robust, li2012feature, pan2019ball, liu2020model, zhong2023feature}, including approaches for false discovery rate control \citep{tong2023model} and threshold selection rules \citep{guo2023threshold}.
For classification problems, several model-free screening methods have been proposed for continuous features \citep{mai2013kolmogorov, mai2015fused, cui2015model, xie2020category,li2024feature}. For categorical features and categorical response, \cite{huang2014feature} proposed a screening procedure based on the Pearson chi-squared statistic (PC-SIS). 
All these methods typically assume independence. Yet in network data, both responses and features may be interconnected within the network. Unlike independent settings, network analysis involves features that also contribute to network formation, as discussed by \cite{yan2019statistical} and \cite{zhang2022joint}. 
Nodal responses are influenced jointly by individual features and the surrounding network.

\begin{figure}[h]
	\centering
    \includegraphics[width=14cm]{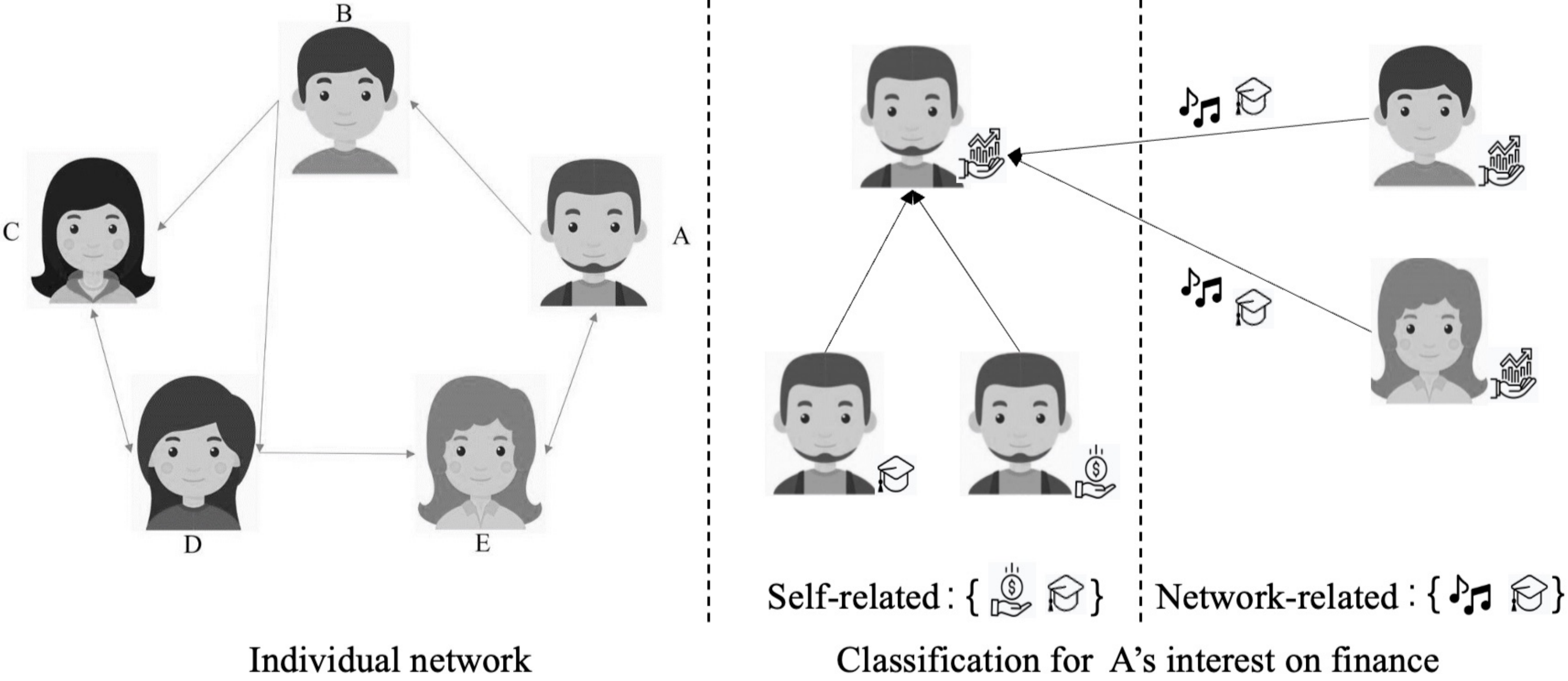}
	\caption{The individual network and network-involved classification  task. The left panel describes the observed network structure among the five nodes.
		The middle panel shows that the response, that is, whether node A is interested in finance, is influenced by node A's own features: income level and education (which are referred to as  {\it  self-related} features).
		The right panel shows that the response of node A could be related to the individual's connected nodes' response and edge generation probabilities between individuals. Hobbies such as music, and education represent {\it network-related} features that contribute to edge generation probability.}\label{intro_fig}
\end{figure}

Consequently, the relationship between features and the response can be categorized into two types: (1) ``{\it self-related}'' features, which directly affect an individual's own response (as in traditional independent settings); and (2) ``{\it network-related}'' features, which influence responses through the network.
{\it Network-related} features first impact the probability of edge formation between nodes (i.e., the edge generation probability), and subsequently influence responses indirectly. Figure \ref{intro_fig} provides an illustrative example involving five individuals.
However, existing methods focus exclusively on {\it self-related} features, neglecting {\it network-related} ones. This limitation becomes especially critical in datasets with network dependencies, where the feature-response relationship is more complex. 
Such oversight may result in incorrect feature screening results and suboptimal predictions. This highlights the necessity of the proposed method in this study.

\subsection{Contributions of This Study}

We focus on an ultra-high-dimensional feature screening problem with categorical response and features for network data, a unique challenge not addressed by previous research. 
We develop a model-free sure independence screening method based on log pseudo-likelihood ratio statistics (PLR-SIS). The statistic incorporates information from network structure, categorical feature, and response. It measures the marginal contribution of each feature without assuming a specific model for the response and the feature. 
The {\it network-related} features contribute to the generation of the network structure, thereby influencing the likelihood of the response. 
In this way, PLR-SIS can identify both {\it self-related} features and {\it network-related} features, both of which contribute to response prediction. 

The contributions of this study are summerized as follows:
\begin{itemize}
    \item First, we conduct an innovative quantitative study of two types of related features—{\it self-related} and {\it network-related}—in the context of feature screening for network data.
    \item Second, we propose a model-free PLR-SIS method designed for categorical responses and features, which is applicable in a variety of scenarios. 
    Numerical analysis using real social network data shows that the proposed method outperforms traditional feature screening methods, delivering significantly better prediction performance with far fewer features.
    \item Lastly, PLR-SIS also allows for sparse network structures and provides strong theoretical guarantees, including the strong consistency, ranking consistency and asymptotic distributions properties.
\end{itemize}

The remainder of this paper is organized as follows. Section 2 introduces the classification framework with the observed network structure. The PLR-SIS procedure, its asymptotic properties, and detailed conditions are established and studied. Simulation analysis is presented in Section 3 to evaluate the empirical performance of the proposed PLR-SIS procedure.
Section 4 provides an application of the proposed PLR-SIS procedure based on Sina Weibo data.
We provide concluding remarks in Section 5.
All the technical proofs are presented in the appendix of the Supplementary Material.

\section{Pseudo-Likelihood Ratio Feature Screening for Network Data}

\subsection{Notations}

First, we introduce basic notations for the response, features, network, and feature set. Define $n$ as the number of nodes in the network. Let $Y_i \in \{1,...,R\}$ be the response preference with $R$ levels, and
$X_i = (X_{i1} ,...,X_{ip}) \in \mR^p$ denote the associated features for node $i\in\{1,..., n\}$. 
Let $X_j=(X_{1j} ,...,X_{nj})^\top$ collect the $j$th feature from all $n$ nodes. For simplicity, we assume that all features $X_j$s ($1\leq j\leq p$) have $K$ levels, that is, $X_{ij}\in\{1,...,K\}$. It is easy to generalize this to $X_j$s with different levels. To describe the network structure between individuals, define an adjacency matrix $A=(a_{i_1i_2})\in \mathbb{R}^{n\times n}$ ($1\leq i_1,i_2\leq n$), where $a_{i_1i_2}=1$ if node $i_1$ follows $i_2$ ($i_1\neq i_2$), which means there is an edge from $i_1$ to $i_2$, and $a_{i_1i_2}=0$ otherwise.
Define $\mS \subset \{1, \dots, p\}$ as the feature set, also referred to as the feature support. 
For example, $\mS_F=\{1,...,p\}$ is the full feature set with all features. Corresponding to the feature set $\mS$, define $X_{i,\mS}=(X_{ij}: j \in \mS) \in \mR^{|\mS|}$ as the feature vector for node $i$, and $X_{\mS}=(X_j: j \in \mS) \in \mR^{n\times|\mS|}$ as the feature matrix.

Next, we introduce the notations for the probabilities of response and features, sample size, and node pairs in the network. 
For any $i_1\in\{1,...,n\}$, define $\pi_{j}^k\!=\! P(X_{i_1j}\!=\!k),
\pi_{y}^r\!=\!P(Y_{i_1}\!=\!r), \pi_{yj}^{rk}\!=\!P(Y_{i_1}\!=\!r, X_{i_1j}\!=\!k),
\pi_{y|j}^{r|k}\!=\!P(Y_{i_1}\!=\!r| X_{i_1j}\!=\!k)$, and $\pi_y^{r_1r_2}\!=\! P(a_{i_1i_2}\!=\!1 | Y_{i_1}\!=\!r_1,Y_{i_2}\!=\!r_2)$. Since we typically consider the marginal relation in screening problems, we further define $\pi^{r_1r_2k_1k_2}_{yj} \mathrel{\!=\!} P\left(a_{i_1i_2} \!=\! 1 \mid Y_{i_1} \!=\! r_1, Y_{i_2} \!=\! r_2, X_{i_1j} \!=\! k_1, X_{i_2j} \!=\! k_2\right)$ to be the edge generation probability based on response and only one feature. 

\begin{figure}[h]
	\centering
	\includegraphics[width=14cm]{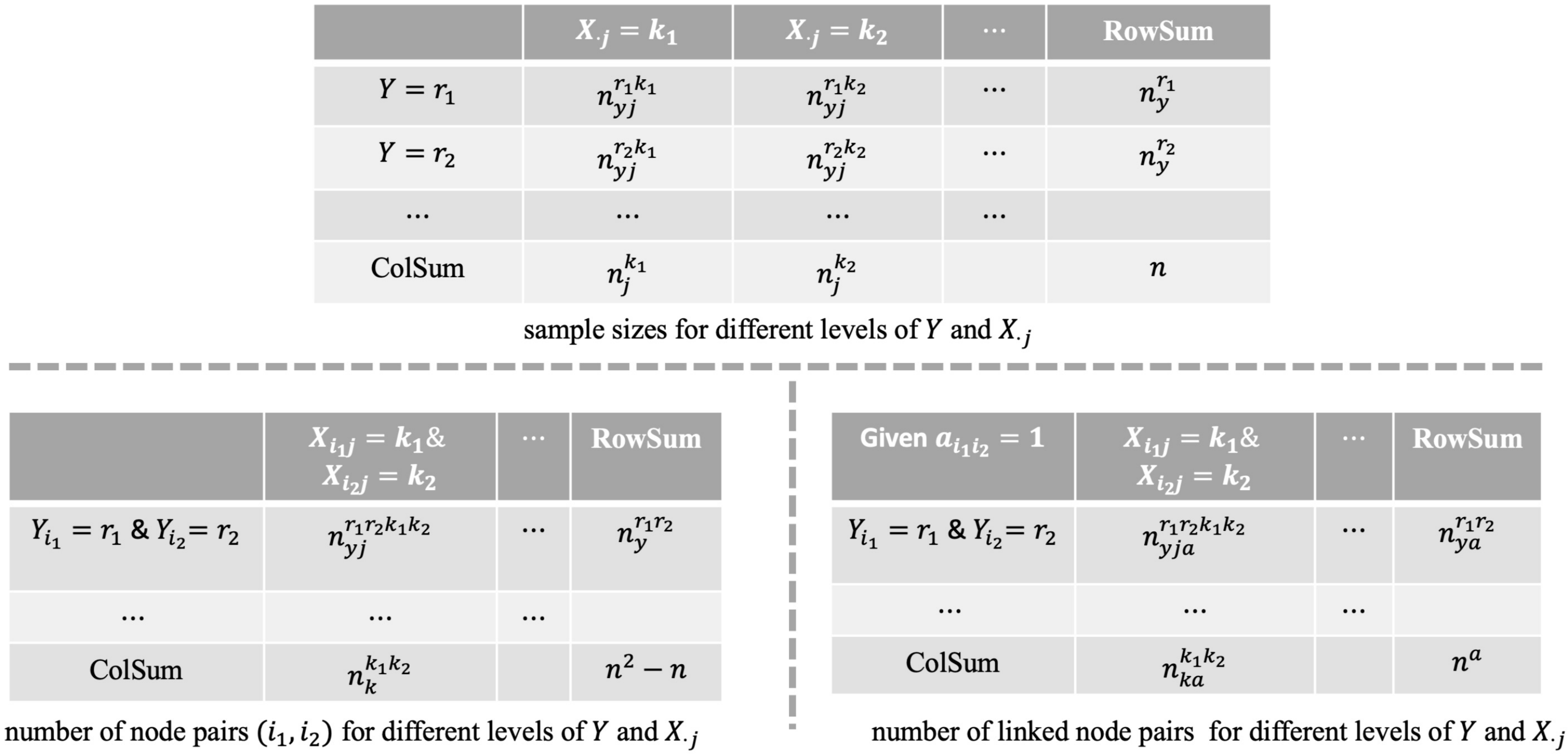}
	\caption{Notations for sample sizes and node pairs. 
    For each table in the figure, ColSum represents the column sum of the table, such as $n_{j}^{k_1}=\sum_{r_1=1}^{R}n_{yj}^{r_1k_1}$. RowSum represents the row sum of the table.}\label{manyn}
\end{figure}

We then introduce the notations for sample sizes under different levels of the response, features, and network connections.
First, we define the sample size of the $r_1$th level of $Y$ as $n^{r_1}_y$. Similarly, the sample size of the $k_1$th level of feature $X_j$ is denoted by $n_j^{k_1}$.
Second, our notation convention is such that subscripts denote variable names (e.g., $y, j$), and superscripts indicate their corresponding  levels (e.g., $r, k$), respectively. For example, $n_{yj}^{r_1k_1} = \sum_i I(Y_i = r_1, X_{ij} = k_1)$ denotes the number of individuals with $Y = r_1$ and $X_j = k_1$.
Further, we define other quantities for node pairs based on the following  convention.
When a superscript contains two levels of the same variable (e.g.,  levels $r_1, r_2$ for $Y$), it refers to the levels associated with a pair of nodes in the network. For instance, $n_y^{r_1r_2}$ denotes the number of node pairs with outcomes $r_1$ and $r_2$, i.e., $n_y^{r_1r_2} \!=\! \sum_{\substack{i_1, i_2 \neq i_1}} I(Y_{i_1} \!=\! r_1, Y_{i_2} \!=\! r_2)$. Further, $n_{yj}^{r_1r_2k_1k_2} \!=\! \sum_{\substack{i_1, i_2 \neq i_1}} I(Y_{i_1} \!=\! r_1, Y_{i_2} \!=\! r_2, X_{i_1j} \!=\! k_1, X_{i_2j} \!=\! k_2)$ denotes the number of node pairs with corresponding levels of $Y$ and $X_j$.
Moreover, a subscript including $a$ (e.g., $n_{yja}^{r_1r_2k_1k_2}$) indicates that the quantity of linked node pairs (i.e., existing edges in the network). For example, 
$n_{yja}^{r_1r_2k_1k_2} \!=\! \sum_{\substack{i_1, i_2 \neq i_1}} a_{i_1i_2} I(Y_{i_1} \!=\! r_1, Y_{i_2} \!=\! r_2, X_{i_1j} \!=\! k_1, X_{i_2j} \!=\! k_2).$
To illustrate our notation rules more intuitively, we summarize the sample size definitions and node pair notations in Figure \ref{manyn}.
\vspace{-0.5cm}
\vspace{-1cm}

\subsection{Data Generation Assumption and Screening Problem}

Following \cite{ma2019flexible}, we adopt a flexible generation framework, where the adjacency matrix is assumed to be generated based on the nodal features and responses. Denote the data generation process
of the network structure is $P(A|X,Y)$,  while $P(X,Y)$ could be modeled
using flexible distributions. For instance, the joint probability
distribution could be expressed as $P(X, A, Y ) = P(A |X , Y )P(Y |X )P(X)$
where we do not need to specify the distribution of $P(X,Y)$. For any node $i$, define $Y_{(-i)}=\{Y_{i_1}:i_1\neq i,,1\leq i_1\leq n\}\in\mR^{n-1}$. Further, we derive the posterior distribution for node-level response prediction $P(Y_i|Y_{(-i)},A)$ and $P(Y_i|Y_{(-i)},A,X_j)$.

Next, we introduce the notations of conditional distribution.
To study the relationships between the response and feature set, we define $\mathcal{D}(Y_i|X_{i,\mS})$ as the conditional distribution of $Y_i$ given $ X_{i,\mS}$. 
Then, we define $\mathcal{D}(a_{i_1i_2}|X_{i_1,\mS},X_{i_2,\mS},Y_{i_1},Y_{i_2})$ as the conditional distribution of $a_{i_1i_2}$ given $ X_{i_1,\mS},X_{i_2,\mS},Y_{i_1},Y_{i_2}$. 
We consider the conditional distribution of $Y_i$ given $Y_{(-i)},A,X_{\mS}$, which is defined as $\mathcal{D}(Y_i|Y_{(-i)}, A, X_{\mS})$.  
Let $\mS_T^*$ be the oracle true feature set, defined as the set of features related to the response. The feature set $\mS_T^*$ is called sufficient, if
\begin{equation}
	\label{suff-dist}
	\mathcal{D}(Y_i|Y_{(-i)}, A, X_{\mS_T^*})=\mathcal{D}(Y_i|Y_{(-i)}, A, X).
\end{equation}
This means that the conditional distribution of the response given the oracle feature support is the same as that given all the features.

However, directly analyzing the sufficiency of $\mS_T^*$ is inconvenient because of the complex expression of the edge generation probability between each pair of nodes. Thus, we adopt the sufficient conditions for equation \eqref{suff-dist}, which are
\beqr
\mathcal{D}(a_{i_1i_2}|X_{i_1,\mS_A},X_{i_2,\mS_A},Y_{i_1},Y_{i_2})&=&\mathcal{D}(a_{i_1i_2}|X_{i_1},X_{i_2},Y_{i_1},Y_{i_2}),\label{suff-dist21}\\
\mathcal{D}(Y_i|Y_{(-i)}, A, X)&=& f\left\{\mathcal{D}(Y_i|X_{i,\mS_Y}),\mathcal{D}(Y_i|A,Y_{(-i)},X_{\mS_A})\right\},\label{suff-dist22}
\eeqr
\noindent
where $f$ is an unknown function, $\mS_A$ is the feature set related to network generation and $\mS_Y$ is related to the generation of the response given the network related information. 
We refer to $\mS_A$ in equation \eqref{suff-dist21} as the feature set of {\it network-related} features and $\mS_Y$ in equation \eqref{suff-dist22}  as the feature set of {\it self-related} features. In this way, by \eqref{suff-dist21}, the marginal distribution of $a_{i_1i_2}$ is determined by response and features in $\mS_A$ of node $i_1$ and $i_2$. 
By \eqref{suff-dist22}, the marginal distribution of $Y_i$ is affected by: {\it  self-related} features $\mS_Y$, observed network structure $A$, other nodal response $Y_{(-i)}$, and {\it network-related} features $\mS_A$. As a result, we define $\mS_T=\mS_A\bigcup\mS_Y$ as the true feature set. It is noteworthy that $\mS_T^* \subset \mS_T$ because \eqref{suff-dist21} and \eqref{suff-dist22} imply \eqref{suff-dist}, making $\mS_T$ a sufficient feature set. However, the identification of $\mS_T$ is much easier than $\mS_T^*$. Thus, empirically, we focus on the identification of $\mS_T$.

Note that conditions \eqref{suff-dist21}–\eqref{suff-dist22} are similar to those in existing research on network models.
For \eqref{suff-dist21}, \cite{yan2019statistical} and \cite{kojevnikov2021limit} argued that the generating mechanism of the network structure is correlated with the nodal covariates, and only a few covariates are significantly correlated. Moreover, \cite{pan2015note} proposed test statistics for \eqref{suff-dist21}. For \eqref{suff-dist22}, \cite{ma2019flexible} and \cite{li2020high} also decomposed the network model into two distinct effects: self-effects and network-effects. This separation makes the model more convenient for optimization and interpretation. In this framework, the feature sets associated with self-effects or network-effects may also differ. Similar model specifications are adopted in \cite{zhang2020logistic} and \cite{zhang2021depth}.

Typically, these two feature sets $\mS_Y$ and $\mS_A$ can overlap
but are not exactly the same.
Obviously, the full model $\mS_F$ is sufficient. Thus, we focus on looking for a sufficient model but with a limited size. The objective here of feature screening is to find a feature set estimator $\widehat{\mS}$ such that: (1) $\mS_T \subset \widehat{\mS}$; and (2) the size of $|\widehat{\mS}|$ is as small as possible.

\subsection{Classification Based on Network Data}
\begin{figure}[h]
	\centering
	\includegraphics[width=14cm]{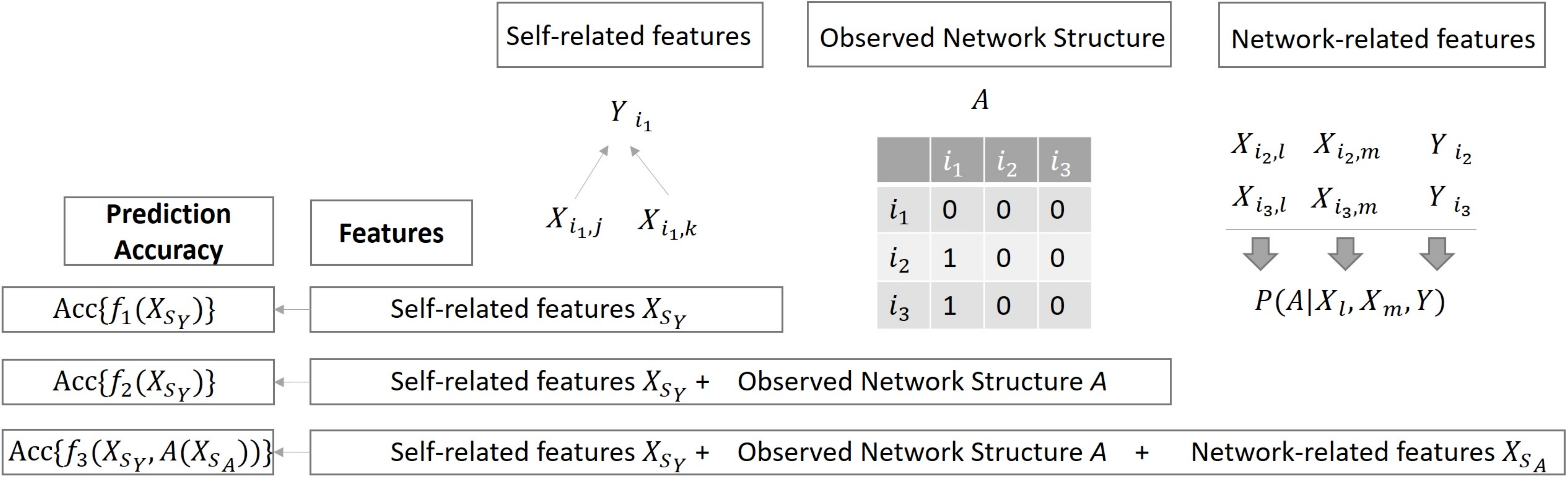}
	\caption{Three types of classifiers and their input information sources. The information sources are: {\it self-related} features $X_{\mS_Y} (\mS_Y=\{j,k\})$; observed network structure $A$; other nodal responses $Y_{(-i)}$; and {\it network-related} features $X_{\mS_A} (\mS_A=\{l,m\})$. }\label{three_predrules}
\end{figure}

Next, we introduce three types of classifiers with different information sources for comparison.
\begin{itemize}
	\item ({\sc Type-I Classifier}). Define the traditional classifier as $f_1(X_{\mS_Y})$, which considers only {\it   self-related} features $\mS_Y$, and does not contain any network information. This is denoted as
	$f_1(X_{\mS_Y}) =\argmax_r P(Y_{i}=r|X_{i,\mS_Y})$.
	\item ({\sc Type-II Classifier}). This classifier considers more than the {\it self-related} features. Considering the observed network structure $A$ and other nodal responses $Y_{(-i)}$, the classifier is denoted as $f_2(X_{\mS_Y})$. This is denoted as $ f_2(X_{\mS_Y})=\argmax_r P(Y_{i}=r|X_{i,\mS_Y},A,Y_{(-i)})$.
	\item ({\sc Type-III Classifier}). Assume we have considered the {\it self-related} features, the observed network structure, and other nodal responses in $ f_3(\mS_Y)$. This classifier further considers the {\it network-related} features, denoted as $f_3(X_{\mS_Y},A(X_{\mS_A}))$, where $\mS_A$ is {\it network-related} features. This can be further denoted as $ f_3(X_{\mS_Y},A(X_{\mS_A}))=\argmax_r P(Y_{i}=r|X_{i,\mS_Y}, A, Y_{(-i)}, X_{\mS_A})$.
\end{itemize}
\noindent See Figure \ref{three_predrules} for illustration. The above classifiers use the maximum posterior probability for node classification. Moreover, the performances of different classifiers are compared by the theoretical prediction accuracy (or Bayes risk, \cite{lugosi2004bayes,james2013introduction,chen2018my}), which is denoted as $\operatorname{Acc}\{f_m(\cdot)\}$ with $m=1,2,3$.

Following the previous literature, we demonstrate the differences in theoretical performances of these three types of classifiers. This leads to the next proposition.

\begin{proposition}
\label{proposition1}
{\it Comparing the prediction accuracy of the Type-I, Type-II, and Type-III classifiers, the following conclusion is obtained:
\begin{equation}
	\label{prop1}
	\operatorname{Acc}\{f_1(X_{\mS_Y})\}\leq \operatorname{Acc}\{f_2(X_{\mS_Y})\}	\leq
	\operatorname{Acc}\{f_3(X_{\mS_Y},A(X_{\mS_A}))\}.
\end{equation} }
\end{proposition}

\noindent 
Detailed proof of Proposition \ref{proposition1} is provided in Appendix A of the Supplementary Material.
The first inequality in \eqref{prop1} illustrates the contribution of the observed network structure and other nodal responses. The performance of the Type-II classifier is typically better than that of Type I. The equality holds when the response 
$Y$ is independent of the network structure 
$A$. The reason is that the observed network structure describes the dependency between nodes. Responses from nodal friends typically help improve the prediction. The second inequality in \eqref{prop1} illustrates the contribution of {\it network-related} features. The involvement of {\it network-related} features in the Type-III classifier can further improve the prediction accuracy, compared with the Type-II classifier.  This is because the {\it network-related} features better describe the probabilities of generating the observed network $A$. 
The equality holds when $\mS_A= \emptyset$ and the network structure is not related to any features. It is important to note that the two conditions for equality to hold are sufficient but not necessary. 
Thus, Proposition \ref{proposition1} indicates that both {\it self-related} features and {\it network-related} features contribute to the response prediction.

In summary, Proposition \ref{proposition1} demonstrates that the Type-III classifier performs best among these three. If we use traditional feature screening methods and ignore the {\it network-related} features, the prediction accuracy of the classifiers may be suboptimal. This leads to the following question of how to obtain such a classifier. Next, we focus on the Type-III classifier and discuss the feature screening method for both {\it self-related} and {\it network-related} features.

\subsection{Definition and Discussion of Pseudo-Likelihood Ratio}

In this section, we assess the impact of a specific feature in Type-III classifiers by comparing posterior probabilities with and without those features.
Ideally, we want to evaluate $P(Y_{i}=r|X_{i,\mS_Y}, A, Y_{(-i)}, X_{\mS_A})$ for each node $i$ ($1\leq i\leq n$). However, in screening problems, we need to consider the marginal relationship between $Y$ and only one feature $X_j$, instead of the feature set $\mS_Y\cup\mS_A$. In this way, we consider the posterior probability with only one feature involved, which is $P(Y_{i}=r|A, Y_{(-i)},X_j)$.
It should be highlighted that $X_j$ may be either {\it self-related} or {\it network-related}. In either possibility, $X_j$ can contribute to the prediction of $Y_{i}$ for node $i$. However, if $X_j$ is not related to $Y$, then whether we include the feature $X_j$ would not make a difference in the prediction of $Y_i$, that is, $\argmax_r P(Y_{i_1}=r|A,Y_{(-i_1)}, X_j) = \argmax_r P(Y_{i_1}=r|A,Y_{(-i_1)}) $. In this way, we consider the differences in the posterior probability of $Y_i$ for two cases: (1) without considering any features; and (2) considering only one feature $X_j$. Consequently, we consider the log pseudo-likelihood ratio based on these two cases. The assumptions made in both cases are solely for constructing the test statistics.

\vspace{0.5cm}
\noindent
\textbf{Case 1. (Without any features)} In this case, we assume the network is determined only by response and is not related to any of the features. Based on the assumption of the classification problem with network structure 
conditional on response, we assume that different edges, $a_{i_1i_2}$, are independent of each other. Thus, $\small {P(A|Y)=\prod_{i_1,i_2\neq i_1}P
 (a_{i_1i_2}|Y_{i_1},Y_{i_2})}$, then we have $P(A|Y)=\prod_{i_1,i_2\neq i_1,r_1,r_2}
\{(\pi_y^{r_1r_2})^{a_{i_1i_2}}(1-\pi_y^{r_1r_2})^{1-a_{i_1i_2}}\}^{I(Y_{i_1}=r_1,Y_{i_2}=r_2)}$.
By the above notations and Bayes' rule, the posterior probability for the response of any node $i_1\in\{1,...,n\}$ is
\begin{equation}
	\begin{aligned}
		P(Y_{i_1}=r_1|A,Y_{(-i_1)})=&	\frac{P(Y_{i_1}=r_1,Y_{{(-i_1)}})P(A|Y_{i_1}=r_1,Y_{(-i_1)})}
		{P(Y_{{(-i_1)}},A)}\\
		\propto  &  \pi_{y}^{r_1} \prod\limits_{i_2\neq i_1,r_2}
		\{(\pi_y^{r_1r_2})^{a_{i_1i_2}}(1-\pi_y^{r_1r_2})^{1-a_{i_1i_2}}\\
		&(\pi_y^{r_2r_1})^{a_{i_2i_1}}(1-\pi_y^{r_2r_1})^{1-a_{i_2i_1}}\}^{I(Y_{i_2}=r_2)}.
		\label{post_prob}
	\end{aligned}
\end{equation}
\noindent where $\propto$ represents ``be proportional to.'' Moreover, the right hand side of ``$\propto$'' in \eqref{post_prob} is denoted as $\ell_{i_1r_1}$.
In other words, the other term not dependent on $r_1$ is omitted.
The detailed calculation is given in Appendix B.1 of the Supplementary Material.
In this case, the prediction rule is written as $\widehat Y_{i_1} = \argmax_{r_1} P(Y_{i_1}=r_1|A,Y_{(-i_1)})=\argmax_{r_1}\ell_{i_1r_1}$.

\vspace{0.5cm}
\noindent
\textbf{Case 2. (With only one feature $X_j$)} In this case, we assume the network is only determined by the response and one feature $X_j$. Similarly, with the assumptions in Case 1, conditional on response and the feature $X_j$, we assume that different edges, $a_{i_1i_2}$, are independent with each other. Thus, we have $P(A|Y,X_j)=\prod_{i_1, i_2\neq i_1}
P(a_{i_1i_2}|Y_{i_1},Y_{i_2},X_{i_1j},X_{i_2j})$.
Then, the posterior probability for the response of any node $i_1\in\{1,...,n\}$ is
\begin{equation}
	\begin{aligned}
		P(Y_{i_1}=r_1|A,Y_{(-i_1)}, X_j)=&P(X_j)P(Y_{i_1}=r_1,Y_{(-i_1)}|X_j)P(A|Y_{i_1}=r_1,\\&Y_{(-i_1)},X_j)/P(A,Y_{(-i_1)},X_j)\\
		\propto&
		\left\{\prod\limits_{k_1= 1}^{K}(\pi_{y|j}^{r_1|k_1})^{I(X_{i_1j}=k_1)}\right\}
		\left[\prod\limits_{i_2\neq i_1,r_2,k_1,k_2}
		\{(\pi_{yj}^{r_1r_2k_1k_2})^{a_{i_1i_2}}\right.\\
		&(1-\pi_{yj}^{r_1r_2k_1k_2})^{1-a_{i_1i_2}}(\pi_{yj}^{r_2r_1k_2k_1})^{a_{i_2i_1}}\\
		&\left.
		(1-\pi_{yj}^{r_2r_1k_2k_1})^{1-a_{i_2i_1}}\}
		^{I(Y_{i_2}=r_2,X_{i_1j}=k_1,X_{i_2j}=k_2)}\right],		
		\label{post_prob2}
	\end{aligned}
\end{equation}
\noindent where the right hand side of ``$\propto$'' in \eqref{post_prob2} is denoted as $\ell^*_{i_1r_1}$. The detailed calculation is given in Appendix B.1 of the Supplementary Material.
In this case, the prediction rule is written as $\widehat Y_{i_1} = \argmax_{r_1} P(Y_{i_1}=r_1|A,Y_{(-i_1)},X_j)=\argmax_{r_1}\ell^*_{i_1r_1}$.

The key difference between \eqref{post_prob} and \eqref{post_prob2} is whether to involve the feature $X_j$ or not.
It may affect the self conditional probability of the response, or affect the network structure's generation.
When $X_j$ is an {\it self-related} feature, $\pi_{y|j}^{r_1|k_1}$ in \eqref{post_prob2} provides more information for prediction than $\pi_{y}^{r_1} $ in \eqref{post_prob}.
When $X_j$ is a {\it network-related} feature, then $\pi_{yj}^{r_1r_2k_1k_2}$ in \eqref{post_prob2} describes the edge generation probability better than $\pi_y^{r_1r_2}$ in \eqref{post_prob}.
Thus, the log ratio of \eqref{post_prob2} and \eqref{post_prob} measures the feature's self-related and network-related effects to the response.

Furthermore, we consider the posterior probability of response for all the nodes.
We consider all the information contributed to nodal response prediction.
Then define $L_{0}= \prod_{i_1,r_1}\ell_{i_1r_1}$ as the pseudo-likelihood without any feature in \textbf{Case 1}.
Further, we calculate the pseudo-likelihood statistic $\widehat L_{0}= \prod_{i_1,r_1}\hat\ell_{i_1r_1}$ as if $Y_{i_1}$s were independent given $A$, and $Y_{(-i_1)}$. In this way, we aim to obtain a simple expression to construct a likelihood-ratio-type statistic.
Similarly define $L_{j}=\prod_{i_1,r_1} \ell_{i_1r_1}^*$ as the pseudo-likelihood with only one feature $X_j$ in \textbf{Case 2}.
Then, we define the log pseudo-likelihood ratio (PLR) as,
\begin{equation}
	\label{LR}
	\Lambda_j=\frac{1}{n}\left(\log L_j-\log L_0\right) .
\end{equation}
\noindent It is remarkable that $\Lambda_j$ measures the marginal contribution of feature $j$ to response prediction. For an irrelevant feature, it is related neither to the response directly nor to the network structure.
Then, for any node $i_1$, the results of \eqref{post_prob} and \eqref{post_prob2} may be similar.
As a result, the log pseudo-likelihood ratio in \eqref{LR} tends to be zero with high probability.
By contrast, for a feature in the true feature set $\mS_T$, the log pseudo-likelihood ratio can be much larger than zero.
We theoretically discuss the log pseudo-likelihood ratios in subsection 2.6. 
Next, we discuss the calculation of \eqref{LR} based on the sample data.

It can be verified that the log pseudo-likelihood ratio can be decomposed into two parts, that is, ${\Lambda}_j = n^{-1}\big( {\Lambda}_j^{\operatorname{self}} + {\Lambda}_j^{\operatorname{network}} \big)$. For the feature $X_j$, the first term ${\Lambda}_j^{\operatorname{self}}$ measures the marginal correlation of the feature with the response, that is, the self-related effect, while ${\Lambda}_j^{\operatorname{network}}$ measures its contribution to the network structure, that is, the network-related effect.
The detailed forms and verification for the decomposition are provided in Appendix B.2 of the Supplementary Material due to the complexity of the symbols.

\subsection{Sure Independence Screening Based on the Pseudo-Likelihood Ratio Statistics}

In this section, we introduce the screening method for the Type-III classifier. 
To estimate $L_0$ and $L_j$, we replace the four types parameters $\pi_{y}^{r_1}$, $\pi_{y|j}^{r_1|k_1}$, $\pi_y^{r_1r_2}$ and $\pi_{yj}^{r_1r_2k_1k_2}$ with their pseudo maximum likelihood estimators (PMLE), respectively. The PMLEs can be proved to be the observation frequencies; for more details, see Appendix B.1 of the Supplementary Material. 
Then based on the notations, we obtain four types of observed frequencies as $\hat{\pi}_{y}^{r_1}=n_y^{r_1}/n$,
$\hat{\pi}_{y|j}^{r_1|k_1}=n_{yj}^{r_1k_1}/n_{j}^{k_1}$,  $\hat{\pi}_y^{r_1r_2}=n_{ya}^{r_1r_2}/n_y^{r_1r_2}$, and $\hat{\pi}_{yj}^{r_1r_2k_1k_2}=n_{yja}^{r_1r_2k_1k_2}/n_{yj}^{r_1r_2k_1k_2}$.
Then, we plug all the frequencies into \eqref{post_prob} and \eqref{post_prob2}, and the following log pseudo-likelihood ratio statistic is obtained by $\widehat{\Lambda}_j = n^{-1}(\log\widehat L_j-\log \widehat L_0)$, where $\widehat L_j=\prod_{i_1,r_1}\hat\ell_{i_1r_1}^*$ and $\widehat L_0=\prod_{i_1,r_1}\hat\ell_{i_1r_1}$ are the estimators of the pseudo-likelihood functions $L_j$ and $L_0$, respectively.

Next, based on the log pseudo-likelihood ratio statistic, we construct the estimated feature set. By the PLR-SIS procedure, the true feature set can be estimated by $\widehat{\mS}=\{j: \widehat{\Lambda}_j> c_*\}$, where $c_{*} > 0$ is a prespecified constant.
Obviously, the selection of the tuning parameter $c_{*}$ is important. Two types of cut-off rules can be considered. For the hard cut-off rule, assume that only \( o(n) \) predictors are truly associated with the response. Define $d=|\mS_T|$ as the number of features in the true feature set. Then we suggest selecting \( d \) as \( [n / \log n] \) or \( n - 1 \), where \( [a] \) denotes the integer part of \( a \) \citep{fan2008sure,pan2018generic}. For the soft cut-off rule, we adopt the maximum ratio criterion used in \cite{huang2014feature}.
Order the $\widehat{\Lambda}_1, ..., \widehat{\Lambda}_p$, such that $\widehat{\Lambda}_{(1)}\geq...\geq \widehat{\Lambda}_{(p)}$. When $j=|\mS_T|$, then $\widehat{\Lambda}_{(j)}> \hat c_{*}$ with probability tending to 1.
For irrelevant features, we have $\widehat{\Lambda}_{(j+1)}\rightarrow_p 0 $. This makes the ratio $\widehat{\Lambda}_{(j)}/\widehat{\Lambda}_{(j+1)}\rightarrow_p \infty$. 
Thus, we have
$\hat{d} = \argmax_{0\leq j\leq p-1} \widehat{\Lambda}_{(j)}/\widehat{\Lambda}_{(j+1)}$, and obtain the estimated $c_{*}$ by $\hat c_{*}=\widehat{\Lambda}_{(\hat{d}+1)}$.
The detailed algorithm of PLR-SIS is shown in Algorithm \ref{alg1}.  After the feature screening procedure, we can further consider the classification models with the observed network structure \citep{cai2018network,zhang2020logistic,zhang2021depth,mukherjee2021high}. Next, we provide theoretical analysis to evaluate the performance of PLR-SIS. 

\begin{algorithm}
	\caption{PLR-SIS procedure}
	\label{alg1}
	\begin{algorithmic}
		\REQUIRE $Y$, $X$, $A$
		\ENSURE $\widehat{\mS}$
		
		\FOR {$j=1,...,p$}
		\FOR {$1\leq r_1,r_2 \leq R$, $1\leq k_1,k_2 \leq K$}
		\STATE Calculate the pseudo maximum likelihood estimators, that is, $\hat\pi_{y}^{r_1}$, $\hat\pi_{y|j}^{r_1|k_1}$, $\hat\pi_y^{r_1r_2}$, and $\hat\pi_{yj}^{r_1r_2k_1k_2}$, by observed frequencies;
		\ENDFOR
		\FOR {$1\leq i_1\leq n,1\leq r_1\leq R$}
		\STATE Calculate $\hat \ell_{i_1r_1}$ and $\hat \ell_{i_1r_1}^*$ by plugging the above pseudo maximum likelihood estimators into \eqref{post_prob} and \eqref{post_prob2};
		\ENDFOR
		\STATE Calculate the pseudo-likelihood statistics $\widehat L_0=\prod_{i_1,r_1} \hat\ell_{i_1r_1}$ and $\widehat L_j=\prod_{i_1,r_1} \hat\ell_{i_1r_1}^*$;
		\STATE Calculate the log pseudo-likelihood ratio statistic $\widehat{\Lambda}_{j}$ for $X_j$ by plugging $\widehat L_0$ and $\widehat L_j$ into \eqref{LR};
		\ENDFOR
		\STATE Order all the log pseudo-likelihood ratio statistics: $\widehat{\Lambda}_{(1)}\geq...\geq \widehat{\Lambda}_{(p)}$;
		\STATE Calculate the estimated size of $\mS_T$ by $\hat{d}=\argmax_{1\leq j\leq p-1} \widehat{\Lambda}_{(j)}/\widehat{\Lambda}_{(j+1)}$ 
		;
		\STATE Obtain the estimated feature set: $\widehat{\mS}=\{j: \widehat{\Lambda}_j > \hat c_{*}\}$ where $\hat c_{*}$ is set to $\widehat{\Lambda}_{(\hat{d}+1)}$.
		\STATE \textbf{Return} $\widehat{\mS}$
	\end{algorithmic}
\end{algorithm}

\begin{remark}
(Extension to feature screening with continuous features). The proposed PLR-SIS procedure is primarily designed for categorical features. 
While for continuous features, by first discretizing the continuous features to be categorical ones, Algorithm \ref{alg1} can also be applied.
For example, define $z_\alpha$ as the $\alpha$th quantile of a standard normal distribution. Redefine $X^*_{ij}$ based on continuous $X_{ij}$ as follows. Define $X^*_{ij}=1$ if $X_{ij}\leq z_{1/K}$, $X^*_{ij}=k$ if $z_{(k-1)/K}<X_{ij}\leq z_{k/K}$, and $X^*_{ij}=K$ if $X_{ij}> z_{(K-1)/K}$.
It is noteworthy that once the true features are accurately identified, the continuous features could be directly incorporated into the classification problem without discretization. 
\end{remark}

\subsection{Theoretical Properties of PLR-SIS}

In this subsection, we investigate the theoretical properties of the PLR-SIS.
Define the self covariance between the response and the $j$th feature as $\omega_{yj}^{rk}=\cov\{I(Y_i=r),I(X_{ij}=k)\}$.
Assume that the response levels $R$, the feature levels $K$, and the number of features in the true feature set $d$ are fixed.
Then, the following conditions are assumed.

\begin{itemize}
	\item[(C1)] (Marginal Probability) Assume that there are positive constants $0<\pi_{\min}<\pi_{\max}<1$, such that $\pi_{\min}<\pi_y^{r}<\pi_{\max}$, $\pi_{\min}<\operatorname{inf}_{1\leq j\leq p}\pi_j^k
	\leq \operatorname{sup}_{1\leq j\leq p}\pi_j^k <\pi_{\max}$, where $1\leq j\leq p$, $1\leq r\leq R$ and $1\leq k\leq K$.
	
	\item[(C2)] (Edge Generation Probability) Assume that the conditional probabilities $\pi_y^{r_1r_2}$ and $\pi_{yj}^{r_1r_2k_1k_2}$ satisfying $ 0<\kappa_1n^{-\gamma}\leq\operatorname{inf}_{1\leq j\leq p}
	\{ \pi_y^{r_1r_2}, \pi_{yj}^{r_1r_2k_1k_2} \}
	\leq\operatorname{sup}_{1\leq j\leq p}
	\{ \pi_y^{r_1r_2},\pi_{yj}^{r_1r_2k_1k_2} \}\\
	\leq 1- \kappa_2n^{-\gamma}< 1$,
	where $\kappa_1,\kappa_2$ are positive constants, $0< \gamma<1$, $1\leq j\leq p$, and $1\leq r_1,r_2\leq R$, $1\leq k_1,k_2\leq K$.
	
	\item[(C3)] (Signal Separation) Assume $\Lambda_j=0$ for $j\notin \mS_T$. Assume $\max_{rk}(\omega_{yj}^{rk})^2>\omega_d$ or $ \max_{r_1r_2k_1k_2}
    \mid\pi_{yj}^{r_1r_2k_1k_2}-\pi_y^{r_1r_2}\mid / \pi_y^{r_1r_2}>\omega_{\operatorname{net}}$ for each $j\in \mS_T$, where $\omega_d$ and $\omega_{\operatorname{net}}$ are positive constants.
	\item[(C4)] ({Divergence Speed}) Assume $\log p\leq \mu \{n\}^\xi$ for two constants $\mu>0$ and $0< \xi< 1$.
\end{itemize}

\noindent Condition (C1) requires that the proportion of each level of the response and features is neither too small nor too large. A similar condition is presented in the previous literature, such as \cite{cui2015model}. Condition (C2) assumes that the edge generation probability should be bounded away from 0 and 1, with a gap depending on $n$. However, it can converge to 0 or 1 at a rate of $n^{-\gamma}$, allowing for certain network sparsity. This condition is similar to those in many network-based classification models
\cite{cai2018network,zhang2020logistic,zhang2021depth}. Condition (C3) ensures the marginal relationship between the response and each feature in $\mS_T$, with at least one level showing a significant correlation. For {\it self-related} features, the marginal correlation must exceed a positive constant, i.e.,  $\max_{rk}(\omega_{yj}^{rk})^2 > \omega_d$. For {\it network-related} features, the relative difference in edge generation probabilities with and without the feature is bounded by a positive constant, i.e., $\max_{r_1r_2k_1k_2} \left|\pi_{yj}^{r_1r_2k_1k_2} - \pi_y^{r_1r_2}\right| / \pi_y^{r_1r_2} > \omega_{\operatorname{net}}$. Irrelevant features neither contribute to the response nor to edge generation. If $X_j$ is independent of both $A$ and $Y$, it could be verified that $\Lambda_{j}=0$ for $j \notin \mS_T$, indicating no relation to the response or the network structure. Lastly, Condition (C4) permits the feature dimension $p$ to grow exponentially with respect to $n$. Based on these conditions, we have the following theorem.

\begin{theorem}
    \label{thm:consistency}
{\it (Strong Screening Consistency)
Under (C1)-(C3), there exists a positive constant $c_*$ such that
$P(\widehat{\mS}=\mS_T)= 1-\eta\exp(\log p-\mu n),$
where $\eta$ and $\mu$ are positive constants. Thus, under (C4), we obtain $P(\widehat{\mS}=\mS_T)\rightarrow 1$ as $n\rightarrow \infty$.}
\end{theorem}

\noindent The proof of Theorem \ref{thm:consistency} is provided in Appendix C.1 of the Supplementary Material. By Theorem \ref{thm:consistency} we conclude that the estimated feature set $\widehat{\mS}$ is exactly equal to the true feature set $\mS_T$ with probability tending to 1. 
In other words, first, all the {\it self-related} and {\it network-related} features can be selected consistently. The $c_*$ is a constant within the interval $[{\Lambda}_{(d+1)}, {\Lambda}_{(d)})$, while in practical computations, Algorithm 1 determines a specific tuning parameter value $\hat{c}_{*}$ is taken to be the estimated value of the left endpoint of the interval, which is the estimated maximum log pseudo likelihood ratio among the redundant features $\widehat{\Lambda}_{(\hat{d}+1)}$.
For $\widehat\Lambda_j$, we establish its ranking consistency property in the following Corollary \ref{cor:ranking-consistency}.

\begin{corollary}
\label{cor:ranking-consistency} {\it (Ranking Consistency Property)
Under (C1)-(C3), as $n\rightarrow \infty$ we have 
$$P(\max_{j \notin \mS_T} \widehat{\Lambda}_j < \min_{j\in {\mS_T}} \widehat{\Lambda}_j)\rightarrow 1.$$}
\end{corollary}
\noindent The proof of Corollary \ref{cor:ranking-consistency} is provided in Appendix C.2 of the Supplementary Material. It shows that the log pseudo-likelihood ratio statistics of all features in the true feature set are asymptotically larger than that of all redundant features. Thus, this further support the determination of the cutoff threshold $c_*$ in Algorithm \ref{alg1}.

In the preference analysis problem, typically, there are features both {\it self-related} and network-related. In other words, they affect both the response directly and the edge generation probability. If we consider a special case, that all the features in the true feature set are {\it network-related} ($\mS_Y\subset\mS_A$), we can allow a faster divergence speed for the dimension $p$ compared with the sample size $n$.
In this case, we replace Conditions (C3)--(C4) with Conditions (C5)--(C6).

\begin{itemize}
	\item[(C5)] (Signal Separation in Network Generation)
	Assume $\Lambda_{j}=0$ for any $j \notin \mS_T$.
	Moreover, for each $j\in \mS_T$, assume
	$\max_{r_1r_2k_1k_2}\mid\pi_{yj}^{r_1r_2k_1k_2}-\pi_y^{r_1r_2}\mid / \pi_y^{r_1r_2}>\omega_{\operatorname{net}}$, where $\omega_{\operatorname{net}}$ is a positive constant.
	\item[(C6)] ({Divergence Speed}) Assume $ \log p\leq \mu \{n^2\}^\xi$ for constants $\mu>0$ and $0< \xi< 1$.
\end{itemize}

\noindent Condition (C5) indicates that all features in the true feature set contribute to the edge generation probability. Condition (C6) allows $p$ to diverge at an exponentially fast speed in terms of $n^2$. Condition (C6) is much milder than (C4).
Then, we have the following corollary.

\begin{corollary}
\label{cor:consistency-network} {\it (Strong Screening Consistency when $\mS_Y\subset\mS_A$)
Under (C1), (C2), and (C5), positive constant $c_*$ exists such that $ P(\widehat{\mS}=\mS_T)= 1-\eta\exp(\log p-\mu n^2)$, where $\eta$ and $\mu$ are positive constants. Then, under (C6), we have $P(\widehat{\mS}=\mS_T)\rightarrow 1$ as $n\rightarrow \infty$.}
\end{corollary}

\noindent The proof of Corollary \ref{cor:consistency-network} is provided in Appendix C.3 of the Supplementary Material. Corollary \ref{cor:consistency-network} indicates that the identification of {\it network-related} features requires a smaller sample size than for features with only marginal self-correlation, as the network provides more information. Specifically, for {\it self-related} features, the correlation is measured across the $n$ individuals for a feature and the response, whereas for {\it network-related} features, the correlation is measured across the $n(n-1)$ node pairs for a feature and their connectivity. Thus, the correlation for {\it network-related} features is more easily identifiable compared to that for {\it self-related} features.

Theorem \ref{thm:consistency} and Corollary \ref{cor:consistency-network} discuss the strong screening consistency for main effects. However, interaction effects are also important in preference analysis and other fields, such as biology and finance. For the proposed PLR-SIS procedure, it is natural to search for important interactions across the $O(p^2)$ feature pairs. Let $ \mS_T^{\mbox{in}}$ denote the true feature set, which includes both features and interactions related to the response. The features and interactions in $ \mS_T^{\mbox{in}}$ can also be identified simultaneously by Algorithm \ref{alg1}. We then obtain the following theorem.

\begin{theorem}
\label{thm:consistency-inter}
{\it(Strong Screening Consistency for Interactions)
Assume that the interactions satisfy the same conditions as in Theorem \ref{thm:consistency} or Corollary \ref{cor:consistency-network}; then, we have $P(\widehat{\mS}^{\operatorname{in}}
=\mS_T^{\operatorname{in}})\rightarrow 1$ as $n\rightarrow \infty$.}
\end{theorem}

\noindent The proof of Theorem \ref{thm:consistency-inter} is provided in Appendix C.4 of the Supplementary Material.
Theorem \ref{thm:consistency-inter} means the PLR-SIS procedure is natural for the interaction detection.
Furthermore, to save computational cost, we can assume that the true feature set  $ \mS_T^{\mbox{in}}$ contains only interactions within $\mS_T$.

Recall that the log pseudo-likelihood ratio is decomposed as the sum of two parts ${\Lambda}_j= n^{-1}\big( {\Lambda}_j^{\operatorname{self}}+{\Lambda}_j^{\operatorname{network}}\big)$. 
Next, we investigate the asymptotic distributions of $\widehat{\Lambda}_j^{\operatorname{self}}$ and $\widehat{\Lambda}_j^{\operatorname{network}}$.
Recall that the response has $R$ levels, and the feature $X_j$ has $K$ levels. Then, we have the following proposition.	

\begin{proposition}
\label{prop:asy-dist}
{\it (Asymptotic Distributions of $\widehat{\Lambda}_j^{\operatorname{self}}$ and $\widehat{\Lambda}_j^{\operatorname{network}}$)
When the null hypothesis holds, that is, feature $X_j$ is neither {\it self-related} nor network-related, we obtain $2\widehat{\Lambda}_j^{\operatorname{self}}\sim \chi^2_{(R-1)(K-1)}$, and $2\widehat{\Lambda}_j^{\operatorname{network}}\sim \chi^2_{R^2(K^2-1)}$.}
\end{proposition}

\noindent The detailed proof of Proposition \ref{prop:asy-dist} is provided in Appendix C.5 of the Supplementary Material. Proposition \ref{prop:asy-dist} shows that, under the null hypothesis, the statistic $2\widehat{\Lambda}_j^{\operatorname{self}}$ follows an asymptotic chi-square distribution with $(R-1)(K-1)$ degrees of freedom, and $2\widehat{\Lambda}_j^{\operatorname{network}}$ follows an asymptotic chi-square distribution with $R^2(K^2-1)$ degrees of freedom. It is noteworthy that, although we only obtain separate asymptotic distributions for $\widehat{\Lambda}_j^{\operatorname{self}}$ and $2\widehat{\Lambda}_j^{\operatorname{network}}$, the random permutation method can be used empirically to approximate the distribution of $\widehat\Lambda_j$ \citep{berrett2020conditional,zhang2024projective}. Therefore, for features with different levels, we do not use $\Lambda_j$ directly due to the differing degrees of freedom across $j$. Instead, we employ the $p$-values obtained through numerical methods for feature screening.

\section{Simulation Studies}
\subsection{Simulation Settings}

To evaluate the finite sample performance of PLR-SIS, we present the following nine simulation examples. We set $p=1000, 2000$, $n=300, 500$, $Y_i\in\{0,1\}$  and $X_{ij}\in\{0,1\}$ in examples 1--5 and 9, $X_{ij}\in\{0,1,2,3\}$ in example 6, $Y_{i}\in\{0,1,2,3\}$  in example 7 and $X_{ij}\in \mR$  in example 8. Below, we illustrate the nine different examples. See Table \ref{simu-case}-Table \ref{simu-case2} for detailed parameters values for each example.

\textbf{Example 1 (Disjoint).}
In this example, we set two disjoint sets $\mS_Y=\{1,2\}$, $\mS_A=\{3,4\}$. In other words, $\mS_Y \cap \mS_A=\emptyset$. Only the $1$st and $2$nd features are self-related, and only the $3$rd and $4$th features are network-related.

\textbf{Example 2 (Network-related).}
In this example, we set $\mS_Y=\{1,2\},\mS_A=\{1,2,3,4\}$. In other words, all the features in the true feature set are {\it network-related} ($\mS_Y\subset\mS_A$). The model generation settings are the same as in Example 1. The key difference is that the $1$st and $2$nd features are not only self-related,
but also network-related.

\textbf{Example 3 (With interaction).} In this example, we set
$\mS_Y=\{1,1\&2\}$, $\mS_A=\{3,4,3\&4\}$, where $1\&2$ and $3\&4$ represent the interaction of the $1$st and $2$nd features and the $3$rd and $4$th features, respectively. The main effect $X_2$ is not significant, while its interaction with the $1$st feature $X_{1\&2}$ is significantly related to the response.

\begin{table}[!h]
\tabcolsep 6pt 
	\caption{Probability specification for Examples 1--3. The abbreviations NNB, NLR, and NET represent network na{\"i}ve Bayes model, network logistic regression model, and the network generation probability, respectively.
	Additionally, for the NNB model in Example 3, the $1$st and $3$rd features are generated by the same conditional probability in Examples 1 and 2.}
	\vspace*{-6pt}
 \renewcommand{\arraystretch}{1.5}
	\begin{center}
		\resizebox{0.6\columnwidth}{!}{ 
			\begin{tabular}
			{@{\hspace{1cm}}cccccc}
   \hline
   Model & Generation probability& $j=1$&$j=2$&$j=3$&$j=4$\\
   \hline
			&\multicolumn {5} {c} {Example 1}\\
			NNB&$P(X_j=1|Y=0)$&0.2&0.9&0.4&0.5 \\
			&$P(X_j=1|Y=1)$ &0.9&0.4&0.4&0.5 \\
			NLR&\multicolumn {5} {l} {$P(Y|X) = \operatorname{Sig}(-4X_1+4X_2)$}\\
			NET&\multicolumn {5} {l} {$P(a_{i_1i_2}|Y,X_{\mS_A})=\operatorname{Sig}( 0.4\sum_{j\in\{3,4\}} S_{i_1i_2,X_j}+\omega_{i_1i_2,Y})$}\\
			\hline
			&\multicolumn {5} {c} {Example 2}\\
			NNB&$P(X_j=1|Y=0)$&0.2&0.9&0.4&0.5 \\
			&$P(X_j=1|Y=1)$ &0.9&0.4&0.4&0.5 \\
			NLR&\multicolumn {5} {l} {$P(Y|X) = \operatorname{Sig}(-4X_1+4X_2)$}\\
			NET&\multicolumn {5} {l} {$P(a_{i_1i_2}|Y,X_{\mS_A})=\operatorname{Sig}(
				0.4\sum_{j\in\{1,2,3,4\}} S_{i_1i_2,X_j}+\omega_{i_1i_2,Y})$}\\
			\hline
			&\multicolumn {5} {c} {Example 3}\\
			NNB&$P(X_{j}=1|Y=1,X_{j-1}=1)$&-- &0.7 &-- &0.8\\
			&$P(X_{j}=1|Y=1,X_{j-1}=0)$&-- &0.2 &-- &0.2\\
			&$P(X_{j}=1|Y=0,X_{j-1}=1)$&-- &0.5 &-- &0.9\\
			&$P(X_{j}=1|Y=0,X_{j-1}=0)$&-- &0.5 &-- &0.1\\
			NLR&\multicolumn {5} {l} {$P(Y|X) = \operatorname{Sig}(3X_1+4X_1X_2)$}\\
			NET&\multicolumn {5} {l} {$P(a_{i_1i_2}|Y,X_{\mS_A})=\operatorname{Sig}(
				0.2\sum_{j\in\{3,4,3\&4\}} S_{i_1i_2,X_j}+\omega_{i_1i_2,Y})$}\\
    \hline
		\end{tabular}}
	\end{center}
	\label{simu-case}
\end{table}

\textbf{Example 4 (Correlated features)} This example introduces correlations between true and redundant features. We set two disjoint feature sets, $\mS_Y=\{1,2\}$, $\mS_A=\{3,4\}$. Additionally, we set $X_{5}$ is redundant feature but is correlated with the {\it self-related} feature $X_{2}$, and $X_{6}$ is another redundant feature correlated with {\it network-related} feature $X_{3}$.

\textbf{Example 5 (Model misspecification)} In this example, we consider network structure observation errors, following the simulation setup by \cite{lewbel2021social}. The observed network structure used for feature screening is given by  $a^{(obs)}_{i_1i_2} = a_{i_1i_2}e_1+(1-a_{i_1i_2})e_2$, where $a_{i_1i_2}$ is the generated true network structure, $e_1\sim \text{Bernoulli}(p_{e_1})$, $e_2\sim \text{Bernoulli}(p_{e_2})$.

\textbf{Example 6 (Multi-level features)} In this example, we set four levels $X_{ij}\in\{0,1,2,3\}$ for each feature $j\in\{1,...,p\}$, with the true feature set $\mS_Y=\{1,2\}$ and $\mS_A=\{3,4\}$.

\textbf{Example 7 (Multi-level response)} In this example, we set four levels response $Y\in\{0,1,2,3\}$, and $X_{ij}\in\{0,1\}$ for $j\in\{1,...,p\}$, with $\mS_Y=\{1,2\}$ and $\mS_A=\{3,4\}$.

\textbf{Example 8 (Continuous features)} In this example, we set $X_{ij}\in \mR$ for each feature $j\in\{1,...,p\}$, with the true feature set $\mS_Y=\{5\}$ and $\mS_A=\{6\}$.

\textbf{Example 9 (Correlation between $\mS_Y$ and $\mS_A$)} In this example, we set two disjoint sets $\mS_Y=\{1,2\}$, $\mS_A=\{3,4\}$. All features except $X_3$ are generated from a Bernoulli distribution with probability 0.5. To introduce  correlation between feature $X_1$ and $X_3$, we specify the conditional distribution of $X_3$ given $X_1$  in Table \ref{simu-case2}.

\textbf{Generation of response and features.} In the simulation examples, we consider the generation of the response $Y$ and features $X$ by following two typical models.

\textbf{Model I}. Network na{\"i}ve Bayes model \cite{huang2018network}. The model is defined as
\begin{equation}
	\begin{aligned}
	\label{nnb}
	\mathrm{P}\left(Y, X, A\right)=
	\left\{\prod_{i} \mathrm{P}\left(Y_{i}\right)\right\}
	\left\{\prod_{i,j} \mathrm{P}\left(X_{i j} \mid Y_{i}\right)\right\}\left\{\prod_{i_{1} \neq i_{2}} \mathrm{P}\left(a_{i_{1} i_{2}} \mid Y_{i_{1}}, Y_{i_{2}},X_{i_1},X_{i_2}\right)\right\},
	\end{aligned}
\end{equation}
where $\mathrm{P}\left(X_{i} \mid Y_{i}=r\right)=\prod_{j\in\mS_Y}(\pi_{j|y}^{1|r})^{X_{i j}}(1-\pi_{j|y}^{1|r})^{1-X_{i j}}$ and $ \pi_{j|y}^{1|r}=P(X_{i}=1| Y_{i}=r)$. By network na{\"i}ve Bayes model \eqref{nnb}, we first generate the response, then the different features.
The response of each node is generated by $P(Y=r) =1/2$ for $r=0,1$. Then, each feature is generated by conditional probability $P(X_j|Y)$; see Table \ref{simu-case}--\ref{simu-case2}.
The feature $X_j$ is {\it self-related} if and only if $P(X_j|Y=0)\neq P(X_j|Y=1)$.
The irrelevant features are generated from a Bernoulli distribution with a probability of 0.2.

\textbf{Model II}. Network logistic regression model \cite{zhang2020logistic}. The model is defined as
\begin{equation}
		\begin{aligned}
	\label{nlr}
	\mathrm{P}\left(Y, X, A\right)=
	\left\{\prod_{i,j} \mathrm{P}\left(X_{ij}\right)\right\}
	\left\{\prod_{i} \mathrm{P}\left(Y_i \mid X_{ij}\right)\right\}
	\left\{\prod_{i_{1} \neq i_{2}} \mathrm{P}\left(a_{i_{1} i_{2}} \mid Y_{i_{1}}, Y_{i_{2}},X_{i_1},X_{i_2}\right)\right\},
		\end{aligned}
\end{equation}
where $\mathrm{P}\left(Y_i \mid X_{i}\right)=\operatorname{Sigmoid}({ X^{\top}_{i,\mS_Y}}\beta)$,
$X_{i,\mS_Y}=(X_{ij}:j\in\mS_Y)\in \mR^{|\mS_Y|}$ is the {\it self-related} feature vector for node $i$, $\operatorname{Sigmoid}(x)
= \exp(x)/\Big\{1+\exp(x)\Big\}$ and $\beta\in\mR^{|\mS_Y|}$ is the regression parameter vector.
By network logistic regression model \eqref{nlr}, we first generate the features, then generate the response.
First, each feature is generated from a Bernoulli distribution with a probability 0.5. Then, the response generation probability is determined by a linear combination of the generated features and their interactions; see details in Table \ref{simu-case}--\ref{simu-case2} for the detailed parameter values.

\textbf{Generation of network structure.} Next, we consider the generation of network structure based on $Y$ and $X$. The entries of the adjacency matrix are independently generated from a Bernoulli distribution with probability $P(a_{i_1i_2}|Y_{i_1},Y_{i_2},S_{i_1i_2,\mS_A})$; see Table \ref{simu-case}.
Typically, we assume that $P(a_{i_1i_2}|Y_{i_1},Y_{i_2},S_{i_1i_2,X_{\mS_A}})=
\operatorname{Sigmoid}( S_{i_1i_2,X_{\mS_A}}^{\top} \phi+\omega_{i_1i_2,Y})$, where $S_{i_1i_2,X_{\mS_A}}=(S_{i_1i_2,j}: j\in\mS_A) \in \mR^{|\mS_A|}$ measures the similarity of the {\it network-related} features between the nodes $i_1$ and $i_2$. Here, we let $S_{i_1i_2,j} = I(X_{i_1j}=X_{i_2j})$. The network parameter $\phi\in\mR^{|\mS_A|}$ is the strength parameter that determines the similarity of the features in the edge generation probability,
and $\omega_{i_1i_2,Y}$ denotes the effect of the response on the edge generation probability. Similar edge generation mechanisms are adopted in \cite{yan2019statistical}, \cite{zhang2020logistic}, and \cite{kojevnikov2021limit}. Moreover, we set the network sparsity parameter $\gamma=0.5$. Define the neighbors' response effect on network generation $\omega_{i_1i_2,Y}=\log(1.0)-\gamma \log(n)$ if $Y_{i_1}=Y_{i_2}$, and $\omega_{i_1i_2,Y}=\log(0.5)-\gamma \log(n)$ otherwise.

\begin{table}[!h]
\tabcolsep 6pt 
	\caption{ Probability specification for Examples 4--9. For Examples 4-8, we utilize the NNB model with network generation identical to Example 1. In Example 5, the feature generation procedure mirrors that of Example 1. In Example 6, the notation $Z$ represents the normalization constant $\sum_k exp(k)$. In Example 9, we introduce a high correlation between $X_1$ and $X_3$, and utilize the NLR model with network generation probability mirrors that of Example 1.}
	\vspace*{-6pt}
 \renewcommand{\arraystretch}{1.5}
	\begin{center}
		\resizebox{\columnwidth}{!}{ 
			\begin{tabular}
			{@{\hspace{1cm}}cccccccc}
   \hline
   Generation probability& $j=1$&$j=2$&$j=3$&$j=4$
   &Generation probability&$j=5$&$j=6$\\
   \hline
	&\multicolumn {5} {c} {Example 4}\\
			$P(X_j=1|Y=0)$&0.3&0.8&0.5&0.6 &$P(X_j=1|X_{j-3}=0)$&0.7&0.8\\
			$P(X_j=1|Y=1)$ &0.9&0.3&0.5&0.6 &$P(X_j=1|X_{j-3}=1)$&0.3&0.2\\
			\hline
			&\multicolumn {5} {c} {Example 5}\\
			
			\multicolumn {8} {c} {$a^{(obs)}_{i_1i_2} = a_{i_1i_2} e_1+(1-a_{i_1i_2})e_2$, $e_1\sim \text{Bernoulli}(1 - n^{s-1})$, $e_2\sim \text{Bernoulli}(10n^{s-2})$, $s=0.4$}\\
			\hline
	&\multicolumn {5} {c} {Example 6}\\
			$P(X_{j}=k|Y=1)$&$e^k/Z$ &0.25 &0.25&0.25 & \multicolumn {3} {c} {mirrors that of Example 1}\\
			$P(X_{j}=k|Y=0)$&0.25 &$e^k/Z$ &0.25 &0.25 & \multicolumn {3} {c} {mirrors that of Example 1}\\
			\hline
	&\multicolumn {5} {c} {Example 7}\\
			$P(X_{j}=1|Y=r)$&$e^{2k}/Z$ &$1/Z$ &0.5&0.5
            &$P(X_{j}=1|Y=r)$& 0.5&0.5\\
			$P(X_{j}=0|Y=r)$&$1/Z$ &$e^{2k}/Z$ &0.5 &0.5 
            &$P(X_{j}=1|Y=r)$&0.5&0.5\\
    \hline
	Generation probability&\multicolumn {4} {c} {Example 9}& Generation probability&\multicolumn {2} {c} {Example 8}\\
			\multicolumn{5}{c}{$ P(X_3\!=\!1|X_1)\!=\!0.5(1+X_1),P(X_3\!=\!0|X_1)\!=\!0.5(1-X_1)$}
            &$P(X_j|Y=1)$& $N(1,1/2)$ &$N(0,1)$\\
			\multicolumn{5}{c}{$P(Y|X)=\operatorname{Sig}(-4X_1+4X_2)$}
            &$P(X_j|Y=0)$&$N(-1,1/2)$&$N(0,1)$\\
    \hline
		\end{tabular}}
	\end{center}
	\label{simu-case2}
\end{table}
\vspace{-0.5cm}

\subsection{Performance Measurements and Simulation Results}

We compare the proposed PLR-SIS with PC-SIS \cite{huang2014feature}, which selects only the {\it self-related} features.
We repeat the experiment for $M=100$ times. Define $\widehat{\mS}^{(m)}$ as the estimated feature set in the $m$th replication.
We adopt four measurements to compare the performances. The first is the average number of correctly identified features $\operatorname{CMF} = M^{-1}\sum_{m=1}^{M} |\mS_T\bigcap \widehat{\mS}^{(m)}|$.
The second is the average number of incorrectly identified features $\operatorname{IMF} =M^{-1}\sum_{m=1}^{M} |\mS_T^c\bigcap \widehat{\mS}^{(m)}|$, where $\mS_T^c=\mS_F\setminus \mS_T$.
The third is the coverage percentage for the $j$th feature $j\in\mS_T$, calculated by $\operatorname{CP}(X_j)=M^{-1}\sum_{m=1}^{M}
I(j\in\widehat{\mS})$.
The last is the average prediction accuracy $\operatorname{Acc}=M^{-1}\sum_{m=1}^{M} P(Y=f(\widehat\mS^{(m)}))$, where $f(\widehat\mS^{(m)})$ is the corresponding maximum posterior probability classifier, concerning PLR-SIS or PC-SIS. By PLR-SIS, the prediction accuracy is obtained by the Type-III classifier. Meanwhile, by PC-SIS, the prediction accuracy is obtained by the Type-I classifier, since it treats individuals as independent.
The detailed simulation results for Example 1 are summarized in Table \ref{simu-res1}.
Results for Examples 2 and 3 are provided in Tables D1 and D2 in Appendix D of the Supplementary Material.

\begin{table}[htbp!]
\tabcolsep 6pt 
	\caption{Simulation results with 100 replications for Example 1 ($\mS_Y\cap\mS_A=\emptyset$).
	The abbreviations CMF, IMF, $\operatorname{CP}(X_j)$, and Acc represent the average number of correctly identified features, the average number of incorrectly identified features, coverage percentage for the $j$th feature $j\in\mS_T$, average prediction accuracy of the corresponding maximum posterior probability classifier, respectively.}
	\vspace*{-6pt}
 \renewcommand{\arraystretch}{1.5}
	\begin{center}
		\resizebox{0.7\columnwidth}{!}{ 
			\begin{tabular}
			{@{\extracolsep{\fill}}ccccr|ccccc}
      \hline
			&&& \multicolumn {6} {l} 
   {Case I: network na{\"i}ve Bayes model }\\
			$p$&$n$&Method&CMF&IMF
			&$\operatorname{CP}(X_1)$&$\operatorname{CP}(X_2)$
			&$\operatorname{CP}(X_3)$&$\operatorname{CP}(X_4)$
			&Acc\\
			\hline
			$1000$&300&PLR-SIS&\bf3.99&\bf0.00 &\bf1.00&\bf0.99&\bf1.00&\bf1.00 &\bf0.98\\
			&&PC-SIS&1.99&0.00&1.00&0.99&0.00&0.00&0.85\\
			&500&PLR-SIS&\bf4.00&\bf0.00&  \bf1.00&\bf1.00&\bf1.00&\bf1.00&\bf0.98\\
			&&PC-SIS&2.00&0.00 &1.00&1.00&0.00&0.00&0.84\\
			\hline
			$2000$&300&PLR-SIS&\bf3.98&\bf0.00 &\bf1.00&0.98&\bf1.00&\bf1.00&\bf0.98\\
			&&PC-SIS&1.99&0.00&1.00&\bf0.99&0.00&0.00&0.84\\
			&500&PLR-SIS&\bf4.00&\bf0.00&  \bf1.00&\bf1.00&\bf1.00&\bf1.00&\bf0.98\\
			&&PC-SIS&2.00&0.00 &1.00&1.00&0.00&0.00&0.84\\
			\hline
			&&& \multicolumn {6} {l} {
				 Case II: network logistic regression model }\\
			$1000$&300&PLR-SIS&\bf3.93&\bf0.00 &0.98&0.95&\bf1.00&\bf1.00&\bf0.97\\
			&&PC-SIS&1.98&0.00&\bf1.00&\bf0.98&0.00&0.00&0.74\\
			&500&PLR-SIS&\bf4.00&\bf0.00&  \bf1.00&\bf1.00&\bf1.00&\bf1.00&\bf0.97\\
			&&PC-SIS&2.00&0.00 &1.00&1.00&0.00&0.00&0.75\\
			\hline
			$2000$&300&PLR-SIS&\bf3.85&\bf0.00 &0.94&0.91&\bf1.00&\bf1.00&\bf0.96\\
			&&PC-SIS&1.98&0.00&\bf1.00&\bf0.98&0.00&0.00&0.74\\
			&500&PLR-SIS&\bf4.00&\bf0.00&  \bf1.00&\bf1.00&\bf1.00&\bf1.00&\bf0.97\\
			&&PC-SIS&2.00&0.00 &1.00&1.00&0.00&0.00&0.75\\
			\bottomrule
			\label{simu-res1}
		\end{tabular}}
	\end{center}
\end{table}

Based on the simulation results, we draw the following conclusions.
First, as the sample size increases, the average number of correctly identified features, CMF, of PLR-SIS increases toward the true feature set size $d$, and the average number of incorrectly identified features, IMF, remains 0.
This reflects that PLR-SIS can select both {\it self-related} and {\it network-related} features.
By contrast, PC-SIS fails to select the {\it network-related} features.
Second, in Example 2, the findings in Table D1 demonstrate the theoretical results in Corollary \ref{cor:consistency-network}. They reveal that PLR-SIS can identify {\it network-related} features requiring a smaller sample size than {\it self-related} features can.
Third, the findings in Table D2 for Example 3 demonstrate the results in Theorem \ref{thm:consistency-inter}. They show that both features and interactions in the true feature set can be identified by PLR-SIS simultaneously. Finally, in each example, the prediction accuracy of PLR-SIS is much higher than that of PC-SIS.
This is because PC-SIS treats individuals as independent and ignores the {\it network-related} features. Meanwhile, PLR-SIS integrates the {\it network-related} information, which contributes to the response prediction.

Next, we present the results of Examples 4--9 in Table D3 in Appendix D of the Supplementary Material. In the correlated features Example 4, PLR-SIS accurately identifies the true features. Features $X_5$ and $X_6$ are also selected  by only 1\% of the
simulation replications. This is due to their high correlation with the true features, violating condition (C3). 
In the model misspecification Example 5, with network structure observation errors introduced and $s=0.4$ in Table \ref{simu-case2}, PLR-SIS could still select the true features, matching the results in \cite{lewbel2021social}.
In the multi-level features and response Examples 6 and 7, PLR-SIS could identify the true feature set effectively. At the 5\% significance level, empirical coverage probabilities are approximately 95\%, validating Proposition \ref{prop:asy-dist}.
In the continuous features Example 8, PLR-SIS successfully identifies the true feature set using quartile-based binning with $b = 4, 10, 20$, where $b$ represents the number of categories defined by quantiles. Thus, PLR-SIS shows its ability to handle continuous features.
Finally, in Example 9 with correlation between $\mS_Y$ and $\mS_A$, the PLR-SIS achieves higher than 97\% coverage rate in recovering true features, demonstrating robustness against dependencies between $S_Y$ and $S_A$ features.

\section{Real Data Example of Sina Weibo}

\begin{figure}[h]
	\centering
	\includegraphics[width=14cm]{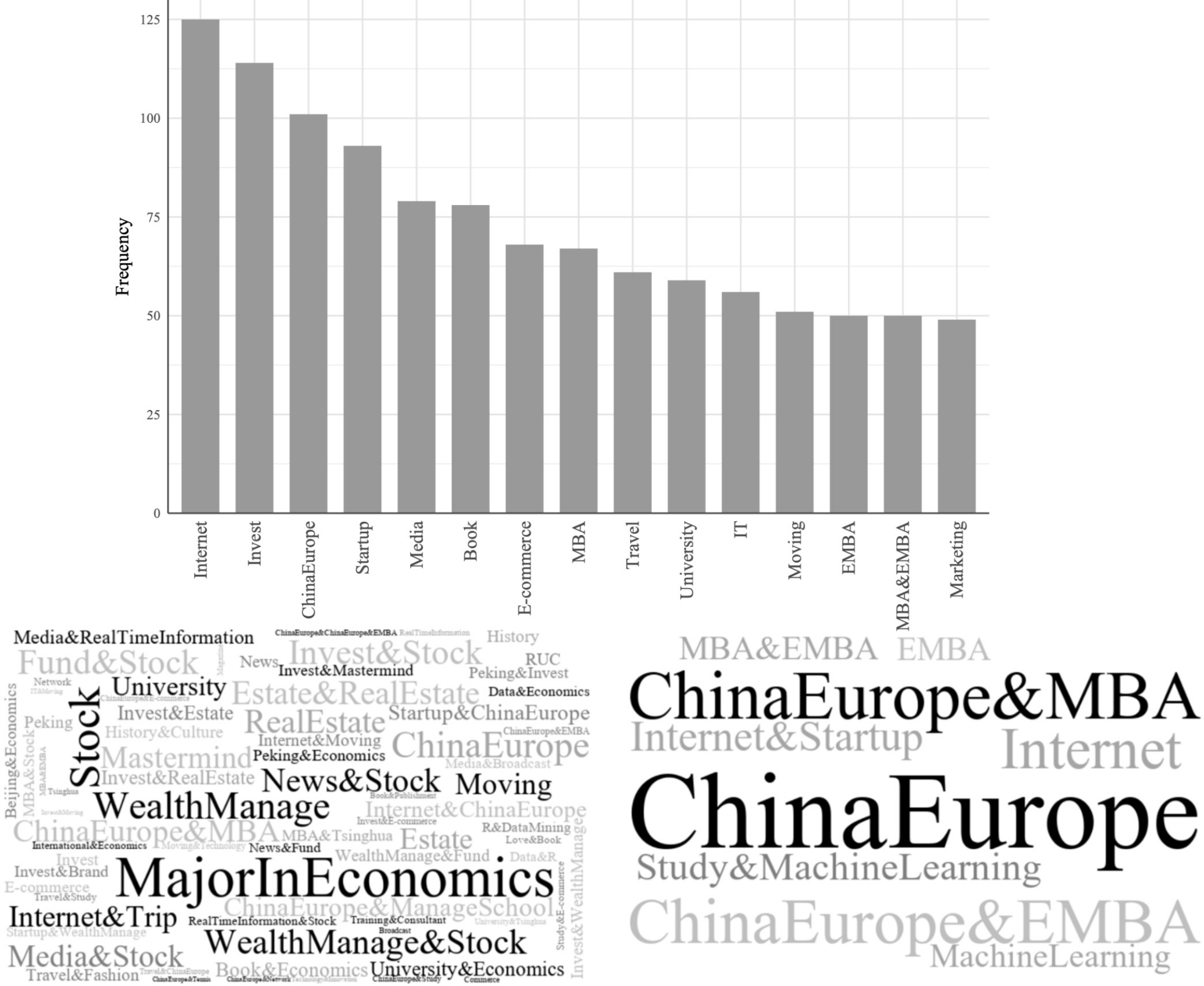}
	\caption{Frequency of the top-15 most common preferences and word cloud diagrams of the screened features.
		The lower left panel corresponds to PC-SIS, while the lower right panel corresponds to PLR-SIS. In each panel, the size of the feature is proportional to the screening statistic.
		The abbreviation ``ChinaEurope'' represents the tag China Europe International Business School.
	}\label{freqwordcloud}
\end{figure}

In this section, we present a real data example from Sina Weibo.
The dataset is collected from the followers of the official account of China Europe International Business School (ChinaEurope), which is a famous business school in China and offers Master of Business Administration (MBA) and Executive Master of Business Administration (EMBA) programs.
The Sina Weibo platform encourages each user to create personalized preferences, such as MBA, food, and music, which may be helpful for understanding users' interests. Moreover, users' informative preferences help explain the network structure.
Users with degrees less than 50 are removed from the dataset.
Then, the dataset includes $n=691$ users and $9870$ features (including 140 tags, and their 9730 interactions between the two tags). 
An interaction term refers to the product of two different features. In the real data analysis, an interaction with a value of 1 indicates that an individual possesses both tags. Including interaction terms allows us to segment the populations.
The bar plot of the top-15 most common preferences is provided in the upper panel of Figure \ref{freqwordcloud}.
The binary response, i.e., the targeted preference, indicates whether the user is interested in finance/economics, with $Y=1$ if the user labeled himself or herself as ``finance/economics''.
The proportion of nodes with $Y=1$ is 39.9\%.
Then we consider applying feature screening. In what follows, we present the results of feature screening, parameter estimation, and prediction performance.

\textbf{Feature screening results.}
Based on the whole dataset, we apply PLR-SIS or PC-SIS to screen the features, which may contribute to the preference analysis. The word cloud diagrams of the selected features are provided in the lower panel of Figure \ref{freqwordcloud}.
First, PLR-SIS selects $|\widehat\mS_Y \cup \widehat\mS_A| = 9$ features. 
These features, "ChinaEurope" and "ChinaEurope\&MBA," belong to the set of important features $S_Y \cap S_A$, which represent the user's affiliation and help explain edge generation in the network. 
This result aligns with the current scenario.
The selected features also reflect the major interests of these users, such as "Machine Learning". The corresponding components $\widehat{\Lambda}^{\operatorname{self}}_j$ and $\widehat{\Lambda}^{\operatorname{network}}_j$ for the features selected by PLR-SIS are illustrated in a stacked bar chart in Figure D1 in Appendix D  of the Supplementary Material, visualizing the marginal contributions of each feature to both the estimated self-related and network-related effect.
In contrast, the number of features selected by PC-SIS is 77, significantly larger than 9 selected by PLR-SIS. The features selected by PC-SIS include broader interests, such as "Stock," "Wealth Management," and "Estate." Meanwhile, the screening statistic value for the feature "ChinaEurope\&MBA" is 11.62, ranking 12th, which is relatively low in this case. This suggests that PC-SIS selects many redundant features, as expected, since it does not consider network structure information.

\begin{figure}[h]
	\centering
	\includegraphics[width=10.5cm]{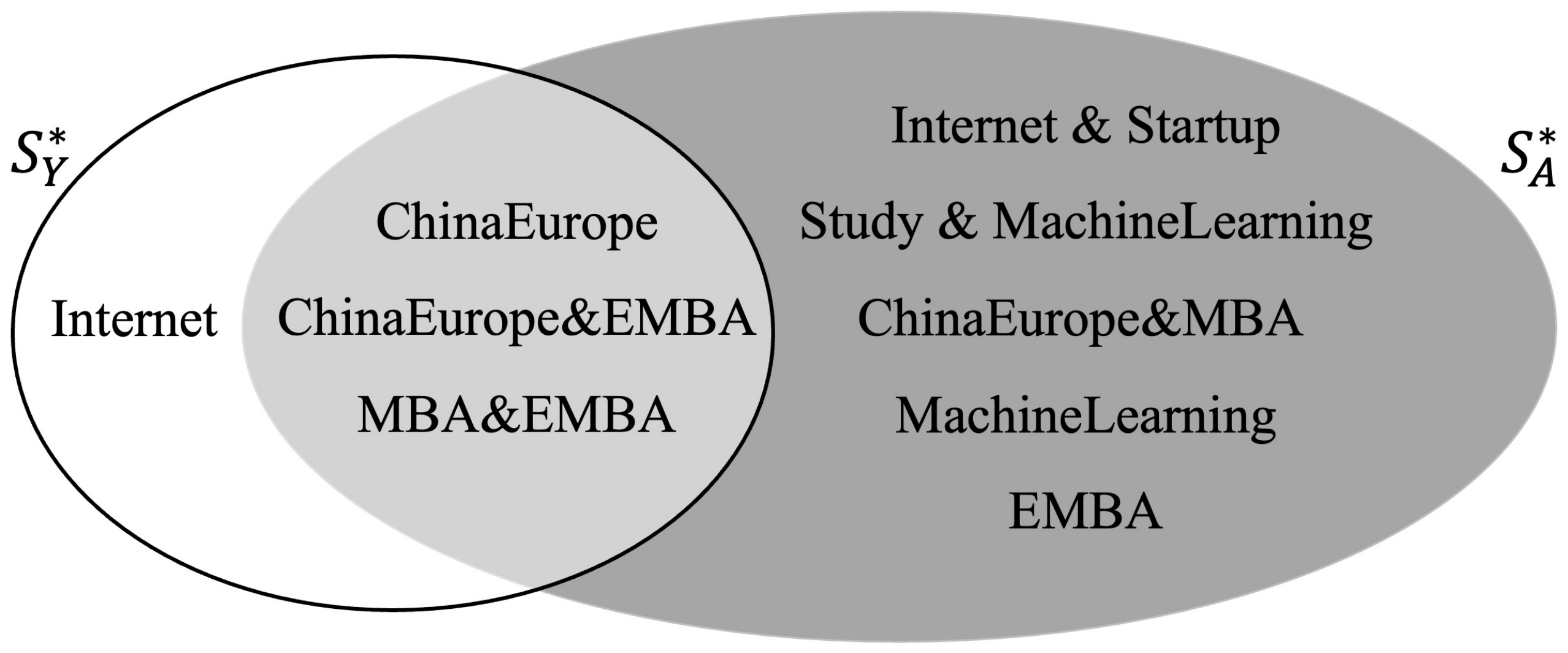}
	\caption{
		Venn diagram of $\mS_Y^*$ and $\mS_A^*$.
		The left ellipse represents $\mS_Y^*$ while the right ellipse represents $\mS_A^*$.
		The left white part, $\mS_Y^*\setminus\mS_A^*$, represents features with significant self-related effects but insignificant network-related effects.
		The middle light gray part, $\mS_Y^*\bigcap \mS_A^*$, represents features with significant self and network-related effects.
		The right dark gray part, $\mS_A^*\setminus\mS_Y^*$, represents features with significant network-related effects but insignificant self-related effects. The notation $\&$ represents the interaction between two tags.
	}\label{vennfeatures}
\end{figure}

\textbf{Parameter estimation results.}
Here, we focus on the detection of significant {\it  self-related} feature set $\mS_Y^*$ and significant {\it network-related} feature set $\mS_A^*$.
In other words, $\mS_Y^*$ is the set of features with significant self-related effects, while
$\mS_A^*$ is the set of features with significant network-related effects.
Based on features selected by PLR-SIS, the estimation results are displayed in Table \ref{real-res1}. Moreover, the Venn diagram of $\mS_Y^*$ and $\mS_A^*$ is intuitively shown in Figure \ref{vennfeatures}.
We can draw the following conclusions.
First, the significance of the estimates shows that $\mS_A^*\setminus\mS_Y^*\neq \emptyset$.
This suggests that some features do not directly affect $Y$, but indirectly affect $Y$ through the network.
Second, for more details, many features selected by PLR-SIS, such as ``ChinaEurope'' and ``ChinaEurope\&EMBA,'' are related to the network structure. This is as expected.
These features are both {\it network-related} and {\it self-related}.
Moreover, such features as ``Machine Learning'' and ``Internet\&Startup'' affect the generation of the network structure, which are the {\it network-related} features.
The feature ``Internet'' is an {\it  self-related} feature, which means that it affects the response directly, but is not involved in the network generation. 
Notably, “Internet\&Startup” represents a finer-grained subgroup within the broader “Internet” category. However, the overall “Internet” tag does not exhibit a significant effect on network structure, as its influence on edge formation is statistically insignificant. 
This interpretation aligns with real-world scenarios, where fine-grained feature combinations (e.g., “Internet\&Startup”) can reveal network-related patterns that may be obscured when analyzing individual features alone (e.g., “Internet”). 
This reflects the inherent complexity of real-world data, in which features and their interactions may exhibit distinct relationships with both the network and the response variables.

\begin{table}[!h]
	\caption{Parameter estimates with features selected by PLR-SIS. 
	The notation $*$ denotes that the estimate is significant at a level of 0.05.
	The model corresponds to the NLR model in Subsection 3.1, where $P(Y|X) = \operatorname{Sigmoid}(\sum_{j\in\mS_Y}X_j\beta_j)$, and $P(a_{i_1i_2}|Y,X)=\operatorname{Sigmoid}(\omega_{i_1i_2,Y}+
	\sum_{j\in\mS_A}\phi_j S_{i_1i_2,X_j})$. The abbreviations CE and ML represent the features China Europe International Business School and Machine Learning, respectively.
	The abbreviations  M\&EMBA and Int\&Sta represent the interaction of  MBA and EMBA,  and that of Internet and Startup, respectively.}
	\vspace*{-6pt}
 \renewcommand{\arraystretch}{1.5}
	\begin{center}
		\resizebox{0.9\columnwidth}{!}{ 
			\begin{tabular}
			{@{\extracolsep{\fill}}cccccc}
   \toprule[1.5pt]
			Feature name&
			Int\&Sta & Study\&ML & ML & CE\&MBA & EMBA \\
			\hline
			Self-effect ($\hat{\beta}_j$)
			&$0.208$&-$0.251$&$0.009$&-$0.149$
			&$0.009$\\
			Network-effect ($\hat{\phi}_j$)
			&-$0.486^*$&$0.665*$&$0.667^*$&-$0.191^*$
			&$0.274^*$ \\
            Feature type & \multicolumn {5} {c} {$j\in \mS_A^*\& j\notin \mS_Y^*$}\\
	\toprule[1.5pt]
			Feature name&
			Internet& CE\&EMBA & CE&  M\&EMBA  \\
			\hline
			Self-effect ($\hat{\beta}_j$)
			&-$1.925^*$&-$1.605^*$&$-0.835^*$&$1.460^*$\\
			Network-effect ($\hat{\phi}_j$)
			&$0.035$&-$0.921^*$ &$0.778^*$&$0.275^*$\\  
            Feature type & $j\in \mS_Y^*\& j\notin \mS_A^*$&
            \multicolumn {3} {c}{$j\in \mS_Y^*\bigcap\mS_A^*$}\\
            \bottomrule[1.5pt]
			\label{real-res1}
		\end{tabular}}
	\end{center}
\end{table}
\vspace{-1cm}
\begin{table}[!h]
\tabcolsep 6pt 
	\caption{Prediction performances with 50 replications for different feature screening procedures.
		The index $p$-value represents the $p$-values of the paired t-test for AUC values between PC-SIS and  PLR-SIS-II (or PLR-SIS-III).
		While the index $p$-value-II represents that between PLR-SIS-II and PLR-SIS-III.}
	\vspace*{-6pt}
 \renewcommand{\arraystretch}{1.5}
	\begin{center}
		\resizebox{0.6\columnwidth}{!}{ 
			\begin{tabular}
			{@{\extracolsep{\fill}}cccccc}
   \hline
			Index& PC-SIS&  PLR-SIS-II  &PLR-SIS-III\\
			\hline
			Averaged AUC &0.728 & 0.758& 0.765 \\
			 $p$-value  &--&1.248$\times10^{-5}$&2.864$\times10^{-7}$ \\
			 $p$-value-II  &--&--&3.116$\times10^{-16}$ \\
   \hline
			\label{real-res2}
		\end{tabular}}
	\end{center}
\end{table}
\vspace{-0.5cm}

\textbf{Prediction Performance Comparison.} Based on the screened features, we consider the prediction of the Type-I, -II, and -III classifiers with logistic regression.
We consider the Type-I classifier for PC-SIS, since it does not consider network information. Then, we consider the Type-II and -III classifiers with {\it network-related} information involved for PLR-SIS.
For simplicity, we refer to the two results as PLR-SIS-II and PLR-SIS-III, respectively.
To further compare the prediction performances of these three classifiers, we consider $70\%$ of the sample for training by random splitting, and the remainder for testing.
We take the average of the area under curve (AUC) values on the test set with 50 replications.
The differences in the AUC values in the three classifiers are all significant by the paired $t$-test. The results are shown in Table \ref{real-res2}.
We can draw the following conclusions.
First, the averaged AUC value of PLR-SIS-II is higher than that of PC-SIS.
This shows that the observed network structure contributes to the preference analysis.
Second, PLR-SIS-III achieves the best prediction performance.
This demonstrates that the {\it network-related} features also contribute to the preference analysis, which corroborates our conclusion in Proposition \ref{proposition1}.
It is remarkable that, based on far fewer features, PLR-SIS-III performs better than PC-SIS.

\section{Conclusion}
In this study, we developed the PLR-SIS procedure for the classification problems with network structure. We first argued that both {\it self-related} and {\it network-related} features could help in prediction accuracy. To select the two types of features simultaneously, we examined a pseudo-likelihood ratio statistic screening procedure.
Furthermore, we investigated the asymptotic theoretical properties of the proposed PLR-SIS method under different scenarios, and presented a number of numerical studies.
In addition, for the considered preference analysis problem in the Sina Weibo data,
the proposed PLR-SIS procedure outperforms the traditional feature screening method with much better prediction performance and far fewer features.
The model-free PLR-SIS method is adaptable to various network data scenarios. For example, in a citation network, authors are nodes, with co-authorship forming the network structure. Research field keywords are ultra-high dimensional features, and the response could be whether a work is highly cited.  Applying the PLR-SIS method could reveal latent patterns in scientific development.

We conclude by discussing three future research directions. First, we assumed a static network structure, but networks evolve over time, suggesting that time-varying networks for feature screening is an interesting area for future study. 
Second, we considered a finite number of response levels \( R \), but extending feature screening to diverging \( R \) is worth exploring, as discussed in \cite{cui2015model} and \cite{xie2020category}. 
Finally, PLR-SIS treats discretized continuous variables as categorical, which does not fully utilize their ordered information. Developing better statistics to preserve this information is a promising topic for future research.

\begin{acks}[Acknowledgments]
This research was supported by Public Computing Cloud, Renmin University of China.
Danyang Huang and  Bo Zhang are the corresponding authors.
\end{acks}

\begin{funding} 
This research is supported by the National Natural Science Foundation of China (NSFC, 12071477, 71873137, 72471230), the Special Research Assistant Grant Program of the Chinese Academy of Sciences (E4C00013), and the Building World-Class Universities (Disciplines) Project of Renmin University of China.  
\end{funding}

\begin{supplement}
\textbf{Appendix.}  The appendices are moved to the supplementary file, which includes the proof of Proposition \ref{proposition1}, details on the calculation of the PLR statistic, the theoretical properties of PLR-SIS, the tables of simulation results and a figure of real data example.
\end{supplement}



\bibliographystyle{imsart-nameyear} 
\bibliography{ref}

\begin{thebibliography}{59}

\bibitem[\protect\citeauthoryear{Armillotta and
  Fokianos}{2023}]{armillotta2023nonlinear}
\begin{barticle}[author]
\bauthor{\bsnm{Armillotta},~\bfnm{Mirko}\binits{M.}} \AND
  \bauthor{\bsnm{Fokianos},~\bfnm{Konstantinos}\binits{K.}}
(\byear{2023}).
\btitle{Nonlinear network autoregression}.
\bjournal{The Annals of Statistics}
\bvolume{51}
\bpages{2526--2552}.
\end{barticle}
\endbibitem

\bibitem[\protect\citeauthoryear{Berrett et~al.}{2020}]{berrett2020conditional}
\begin{barticle}[author]
\bauthor{\bsnm{Berrett},~\bfnm{Thomas~B}\binits{T.~B.}},
  \bauthor{\bsnm{Wang},~\bfnm{Yi}\binits{Y.}},
  \bauthor{\bsnm{Barber},~\bfnm{Rina~Foygel}\binits{R.~F.}} \AND
  \bauthor{\bsnm{Samworth},~\bfnm{Richard~J}\binits{R.~J.}}
(\byear{2020}).
\btitle{The conditional permutation test for independence while controlling for
  confounders}.
\bjournal{Journal of the Royal Statistical Society Series B: Statistical
  Methodology}
\bvolume{82}
\bpages{175--197}.
\end{barticle}
\endbibitem

\bibitem[\protect\citeauthoryear{Bruggeman}{2013}]{bruggeman2013social}
\begin{bbook}[author]
\bauthor{\bsnm{Bruggeman},~\bfnm{Jeroen}\binits{J.}}
(\byear{2013}).
\btitle{Social networks: An introduction}.
\bpublisher{Routledge}.
\end{bbook}
\endbibitem

\bibitem[\protect\citeauthoryear{Cai et~al.}{2018}]{cai2018network}
\begin{barticle}[author]
\bauthor{\bsnm{Cai},~\bfnm{Wei}\binits{W.}},
  \bauthor{\bsnm{Guan},~\bfnm{Guoyu}\binits{G.}},
  \bauthor{\bsnm{Pan},~\bfnm{Rui}\binits{R.}},
  \bauthor{\bsnm{Zhu},~\bfnm{Xuening}\binits{X.}} \AND
  \bauthor{\bsnm{Wang},~\bfnm{Hansheng}\binits{H.}}
(\byear{2018}).
\btitle{Network linear discriminant analysis}.
\bjournal{Computational Statistics \& Data Analysis}
\bvolume{117}
\bpages{32--44}.
\end{barticle}
\endbibitem

\bibitem[\protect\citeauthoryear{Campbell, Ferraro and
  Sands}{2014}]{campbell2014segmenting}
\begin{barticle}[author]
\bauthor{\bsnm{Campbell},~\bfnm{Colin}\binits{C.}},
  \bauthor{\bsnm{Ferraro},~\bfnm{Carla}\binits{C.}} \AND
  \bauthor{\bsnm{Sands},~\bfnm{Sean}\binits{S.}}
(\byear{2014}).
\btitle{Segmenting consumer reactions to social network marketing}.
\bjournal{European Journal of Marketing}.
\end{barticle}
\endbibitem

\bibitem[\protect\citeauthoryear{Chang and Paul}{2024}]{chang2024embedding}
\begin{barticle}[author]
\bauthor{\bsnm{Chang},~\bfnm{Jae~Ho}\binits{J.~H.}} \AND
  \bauthor{\bsnm{Paul},~\bfnm{Subhadeep}\binits{S.}}
(\byear{2024}).
\btitle{Embedding Network Autoregression for time series analysis and causal
  peer effect inference}.
\bjournal{arXiv preprint arXiv:2406.05944}.
\end{barticle}
\endbibitem

\bibitem[\protect\citeauthoryear{Chen, Johansson and Sontag}{2018}]{chen2018my}
\begin{barticle}[author]
\bauthor{\bsnm{Chen},~\bfnm{Irene}\binits{I.}},
  \bauthor{\bsnm{Johansson},~\bfnm{Fredrik~D}\binits{F.~D.}} \AND
  \bauthor{\bsnm{Sontag},~\bfnm{David}\binits{D.}}
(\byear{2018}).
\btitle{Why is my classifier discriminatory?}
\bjournal{Advances in Neural Information Processing Systems}
\bvolume{31}.
\end{barticle}
\endbibitem

\bibitem[\protect\citeauthoryear{Cui, Li and Zhong}{2015}]{cui2015model}
\begin{barticle}[author]
\bauthor{\bsnm{Cui},~\bfnm{Hengjian}\binits{H.}},
  \bauthor{\bsnm{Li},~\bfnm{Runze}\binits{R.}} \AND
  \bauthor{\bsnm{Zhong},~\bfnm{Wei}\binits{W.}}
(\byear{2015}).
\btitle{Model-free feature screening for ultrahigh dimensional discriminant
  analysis}.
\bjournal{Journal of the American Statistical Association}
\bvolume{110}
\bpages{630--641}.
\end{barticle}
\endbibitem

\bibitem[\protect\citeauthoryear{Fan and Lv}{2008}]{fan2008sure}
\begin{barticle}[author]
\bauthor{\bsnm{Fan},~\bfnm{Jianqing}\binits{J.}} \AND
  \bauthor{\bsnm{Lv},~\bfnm{Jinchi}\binits{J.}}
(\byear{2008}).
\btitle{Sure independence screening for ultrahigh dimensional feature space}.
\bjournal{Journal of the Royal Statistical Society: Series B (Statistical
  Methodology)}
\bvolume{70}
\bpages{849--911}.
\end{barticle}
\endbibitem

\bibitem[\protect\citeauthoryear{Fan, Samworth and Wu}{2009}]{fan2009ultrahigh}
\begin{barticle}[author]
\bauthor{\bsnm{Fan},~\bfnm{Jianqing}\binits{J.}},
  \bauthor{\bsnm{Samworth},~\bfnm{Richard}\binits{R.}} \AND
  \bauthor{\bsnm{Wu},~\bfnm{Yichao}\binits{Y.}}
(\byear{2009}).
\btitle{Ultrahigh dimensional feature selection: beyond the linear model}.
\bjournal{The Journal of Machine Learning Research}
\bvolume{10}
\bpages{2013--2038}.
\end{barticle}
\endbibitem

\bibitem[\protect\citeauthoryear{Foster, West and
  Francescucci}{2011}]{foster2011exploring}
\begin{barticle}[author]
\bauthor{\bsnm{Foster},~\bfnm{Mary}\binits{M.}},
  \bauthor{\bsnm{West},~\bfnm{Bettina}\binits{B.}} \AND
  \bauthor{\bsnm{Francescucci},~\bfnm{Anthony}\binits{A.}}
(\byear{2011}).
\btitle{Exploring social media user segmentation and online brand profiles}.
\bjournal{Journal of Brand Management}
\bvolume{19}
\bpages{4--17}.
\end{barticle}
\endbibitem

\bibitem[\protect\citeauthoryear{Frikha et~al.}{2016}]{frikha2016time}
\begin{binproceedings}[author]
\bauthor{\bsnm{Frikha},~\bfnm{Mohamed}\binits{M.}},
  \bauthor{\bsnm{Mhiri},~\bfnm{Mohamed}\binits{M.}},
  \bauthor{\bsnm{Zarai},~\bfnm{Mounir}\binits{M.}} \AND
  \bauthor{\bsnm{Gargouri},~\bfnm{Faiez}\binits{F.}}
(\byear{2016}).
\btitle{Time-sensitive trust calculation between social network friends for
  personalized recommendation}.
In \bbooktitle{proceedings of the 18th annual international conference on
  electronic commerce: e-commerce in smart connected world}
\bpages{1--8}.
\end{binproceedings}
\endbibitem

\bibitem[\protect\citeauthoryear{Gao et~al.}{2021}]{gao2021cross}
\begin{barticle}[author]
\bauthor{\bsnm{Gao},~\bfnm{Chen}\binits{C.}},
  \bauthor{\bsnm{Lin},~\bfnm{Tzu-Heng}\binits{T.-H.}},
  \bauthor{\bsnm{Li},~\bfnm{Nian}\binits{N.}},
  \bauthor{\bsnm{Jin},~\bfnm{Depeng}\binits{D.}} \AND
  \bauthor{\bsnm{Li},~\bfnm{Yong}\binits{Y.}}
(\byear{2021}).
\btitle{Cross-platform item recommendation for online social e-commerce}.
\bjournal{IEEE Transactions on Knowledge and Data Engineering}
\bvolume{35}
\bpages{1351-1364}.
\bdoi{10.1109/TKDE.2021.3098702}
\end{barticle}
\endbibitem

\bibitem[\protect\citeauthoryear{Guo et~al.}{2023}]{guo2023threshold}
\begin{barticle}[author]
\bauthor{\bsnm{Guo},~\bfnm{Xu}\binits{X.}},
  \bauthor{\bsnm{Ren},~\bfnm{Haojie}\binits{H.}},
  \bauthor{\bsnm{Zou},~\bfnm{Changliang}\binits{C.}} \AND
  \bauthor{\bsnm{Li},~\bfnm{Runze}\binits{R.}}
(\byear{2023}).
\btitle{Threshold selection in feature screening for error rate control}.
\bjournal{Journal of the American Statistical Association}
\bvolume{118}
\bpages{1773--1785}.
\end{barticle}
\endbibitem

\bibitem[\protect\citeauthoryear{Hagiu and Wright}{2020}]{hagiu2020data}
\begin{barticle}[author]
\bauthor{\bsnm{Hagiu},~\bfnm{Andrei}\binits{A.}} \AND
  \bauthor{\bsnm{Wright},~\bfnm{Julian}\binits{J.}}
(\byear{2020}).
\btitle{When data creates competitive advantage}.
\bjournal{Harvard business review}
\bvolume{98}
\bpages{94--101}.
\end{barticle}
\endbibitem

\bibitem[\protect\citeauthoryear{Huang, Li and Wang}{2014}]{huang2014feature}
\begin{barticle}[author]
\bauthor{\bsnm{Huang},~\bfnm{Danyang}\binits{D.}},
  \bauthor{\bsnm{Li},~\bfnm{Runze}\binits{R.}} \AND
  \bauthor{\bsnm{Wang},~\bfnm{Hansheng}\binits{H.}}
(\byear{2014}).
\btitle{Feature screening for ultrahigh dimensional categorical data with
  applications}.
\bjournal{Journal of Business \& Economic Statistics}
\bvolume{32}
\bpages{237--244}.
\end{barticle}
\endbibitem

\bibitem[\protect\citeauthoryear{Huang et~al.}{2018}]{huang2018network}
\begin{barticle}[author]
\bauthor{\bsnm{Huang},~\bfnm{Danyang}\binits{D.}},
  \bauthor{\bsnm{Guan},~\bfnm{Guoyu}\binits{G.}},
  \bauthor{\bsnm{Zhou},~\bfnm{Jing}\binits{J.}} \AND
  \bauthor{\bsnm{Wang},~\bfnm{Hansheng}\binits{H.}}
(\byear{2018}).
\btitle{Network-based naive Bayes model for social network}.
\bjournal{Science China Mathematics}
\bvolume{61}
\bpages{627--640}.
\end{barticle}
\endbibitem

\bibitem[\protect\citeauthoryear{James et~al.}{2013}]{james2013introduction}
\begin{bbook}[author]
\bauthor{\bsnm{James},~\bfnm{Gareth}\binits{G.}},
  \bauthor{\bsnm{Witten},~\bfnm{Daniela}\binits{D.}},
  \bauthor{\bsnm{Hastie},~\bfnm{Trevor}\binits{T.}} \AND
  \bauthor{\bsnm{Tibshirani},~\bfnm{Robert}\binits{R.}}
(\byear{2013}).
\btitle{An introduction to statistical learning}
\bvolume{112}.
\bpublisher{Springer}.
\end{bbook}
\endbibitem

\bibitem[\protect\citeauthoryear{Jiang, Li and
  Yao}{2023}]{jiang2023autoregressive}
\begin{barticle}[author]
\bauthor{\bsnm{Jiang},~\bfnm{Binyan}\binits{B.}},
  \bauthor{\bsnm{Li},~\bfnm{Jialiang}\binits{J.}} \AND
  \bauthor{\bsnm{Yao},~\bfnm{Qiwei}\binits{Q.}}
(\byear{2023}).
\btitle{Autoregressive networks}.
\bjournal{Journal of Machine Learning Research}
\bvolume{24}
\bpages{1--69}.
\end{barticle}
\endbibitem

\bibitem[\protect\citeauthoryear{Jin et~al.}{2021}]{jin2021survey}
\begin{barticle}[author]
\bauthor{\bsnm{Jin},~\bfnm{Di}\binits{D.}},
  \bauthor{\bsnm{Yu},~\bfnm{Zhizhi}\binits{Z.}},
  \bauthor{\bsnm{Jiao},~\bfnm{Pengfei}\binits{P.}},
  \bauthor{\bsnm{Pan},~\bfnm{Shirui}\binits{S.}},
  \bauthor{\bsnm{He},~\bfnm{Dongxiao}\binits{D.}},
  \bauthor{\bsnm{Wu},~\bfnm{Jia}\binits{J.}},
  \bauthor{\bsnm{Yu},~\bfnm{Philip}\binits{P.}} \AND
  \bauthor{\bsnm{Zhang},~\bfnm{Weixiong}\binits{W.}}
(\byear{2021}).
\btitle{A survey of community detection approaches: From statistical modeling
  to deep learning}.
\bjournal{IEEE Transactions on Knowledge and Data Engineering}
\bvolume{35}
\bpages{1149-1170}.
\bdoi{10.1109/TKDE.2021.3104155}
\end{barticle}
\endbibitem

\bibitem[\protect\citeauthoryear{Katona, Zubcsek and
  Sarvary}{2011}]{katona2011network}
\begin{barticle}[author]
\bauthor{\bsnm{Katona},~\bfnm{Zsolt}\binits{Z.}},
  \bauthor{\bsnm{Zubcsek},~\bfnm{Peter~Pal}\binits{P.~P.}} \AND
  \bauthor{\bsnm{Sarvary},~\bfnm{Miklos}\binits{M.}}
(\byear{2011}).
\btitle{Network effects and personal influences: The diffusion of an online
  social network}.
\bjournal{Journal of Marketing Research}
\bvolume{48}
\bpages{425--443}.
\end{barticle}
\endbibitem

\bibitem[\protect\citeauthoryear{Kipf and Welling}{2016}]{kipf2016semi}
\begin{barticle}[author]
\bauthor{\bsnm{Kipf},~\bfnm{Thomas~N}\binits{T.~N.}} \AND
  \bauthor{\bsnm{Welling},~\bfnm{Max}\binits{M.}}
(\byear{2016}).
\btitle{Semi-supervised classification with graph convolutional networks}.
\bjournal{arXiv preprint arXiv:1609.02907}.
\end{barticle}
\endbibitem

\bibitem[\protect\citeauthoryear{Kojevnikov, Marmer and
  Song}{2021}]{kojevnikov2021limit}
\begin{barticle}[author]
\bauthor{\bsnm{Kojevnikov},~\bfnm{Denis}\binits{D.}},
  \bauthor{\bsnm{Marmer},~\bfnm{Vadim}\binits{V.}} \AND
  \bauthor{\bsnm{Song},~\bfnm{Kyungchul}\binits{K.}}
(\byear{2021}).
\btitle{Limit theorems for network dependent random variables}.
\bjournal{Journal of Econometrics}
\bvolume{222}
\bpages{882--908}.
\end{barticle}
\endbibitem

\bibitem[\protect\citeauthoryear{Lewbel et~al.}{2021}]{lewbel2021social}
\begin{bbook}[author]
\bauthor{\bsnm{Lewbel},~\bfnm{Arthur}\binits{A.}},
  \bauthor{\bsnm{Qu},~\bfnm{Xi}\binits{X.}},
  \bauthor{\bsnm{Tang},~\bfnm{Xun}\binits{X.}} \betal{et~al.}
(\byear{2021}).
\btitle{Social Networks with Mismeasured Links}.
\bpublisher{Boston College}.
\end{bbook}
\endbibitem

\bibitem[\protect\citeauthoryear{Li, Levina and Zhu}{2019}]{li2019prediction}
\begin{barticle}[author]
\bauthor{\bsnm{Li},~\bfnm{Tianxi}\binits{T.}},
  \bauthor{\bsnm{Levina},~\bfnm{Elizaveta}\binits{E.}} \AND
  \bauthor{\bsnm{Zhu},~\bfnm{Ji}\binits{J.}}
(\byear{2019}).
\btitle{Prediction models for network-linked data}.
\bjournal{The Annals of Applied Statistics}
\bvolume{13}
\bpages{132--164}.
\end{barticle}
\endbibitem

\bibitem[\protect\citeauthoryear{Li and Xu}{2024}]{li2024feature}
\begin{barticle}[author]
\bauthor{\bsnm{Li},~\bfnm{Xingxiang}\binits{X.}} \AND
  \bauthor{\bsnm{Xu},~\bfnm{Chen}\binits{C.}}
(\byear{2024}).
\btitle{Feature screening with conditional rank utility for big-data
  classification}.
\bjournal{Journal of the American Statistical Association}
\bvolume{119}
\bpages{1385--1395}.
\end{barticle}
\endbibitem

\bibitem[\protect\citeauthoryear{Li, Zhong and Zhu}{2012}]{li2012feature}
\begin{barticle}[author]
\bauthor{\bsnm{Li},~\bfnm{Runze}\binits{R.}},
  \bauthor{\bsnm{Zhong},~\bfnm{Wei}\binits{W.}} \AND
  \bauthor{\bsnm{Zhu},~\bfnm{Liping}\binits{L.}}
(\byear{2012}).
\btitle{Feature screening via distance correlation learning}.
\bjournal{Journal of the American Statistical Association}
\bvolume{107}
\bpages{1129--1139}.
\end{barticle}
\endbibitem

\bibitem[\protect\citeauthoryear{Li et~al.}{2012}]{li2012robust}
\begin{barticle}[author]
\bauthor{\bsnm{Li},~\bfnm{Gaorong}\binits{G.}},
  \bauthor{\bsnm{Peng},~\bfnm{Heng}\binits{H.}},
  \bauthor{\bsnm{Zhang},~\bfnm{Jun}\binits{J.}} \AND
  \bauthor{\bsnm{Zhu},~\bfnm{Lixing}\binits{L.}}
(\byear{2012}).
\btitle{Robust rank correlation based screening}.
\bjournal{The Annals of Statistics}
\bvolume{40}
\bpages{1846--1877}.
\end{barticle}
\endbibitem

\bibitem[\protect\citeauthoryear{Li et~al.}{2020}]{li2020high}
\begin{barticle}[author]
\bauthor{\bsnm{Li},~\bfnm{Tianxi}\binits{T.}},
  \bauthor{\bsnm{Qian},~\bfnm{Cheng}\binits{C.}},
  \bauthor{\bsnm{Levina},~\bfnm{Elizaveta}\binits{E.}} \AND
  \bauthor{\bsnm{Zhu},~\bfnm{Ji}\binits{J.}}
(\byear{2020}).
\btitle{High-dimensional Gaussian graphical models on network-linked data.}
\bjournal{Journal of Machine Learning Research}
\bvolume{21}
\bpages{74--1}.
\end{barticle}
\endbibitem

\bibitem[\protect\citeauthoryear{Liu, Chen and Yang}{2022}]{liu2022prediction}
\begin{barticle}[author]
\bauthor{\bsnm{Liu},~\bfnm{Jie}\binits{J.}},
  \bauthor{\bsnm{Chen},~\bfnm{Haojie}\binits{H.}} \AND
  \bauthor{\bsnm{Yang},~\bfnm{Yang}\binits{Y.}}
(\byear{2022}).
\btitle{Prediction models with graph kernel regularization for network data}.
\bjournal{Journal of Applied Statistics}
\bpages{1--18}.
\end{barticle}
\endbibitem

\bibitem[\protect\citeauthoryear{Liu, Zhong and Li}{2015}]{liu2015selective}
\begin{barticle}[author]
\bauthor{\bsnm{Liu},~\bfnm{JingYuan}\binits{J.}},
  \bauthor{\bsnm{Zhong},~\bfnm{Wei}\binits{W.}} \AND
  \bauthor{\bsnm{Li},~\bfnm{RunZe}\binits{R.}}
(\byear{2015}).
\btitle{A selective overview of feature screening for ultrahigh-dimensional
  data}.
\bjournal{Science China Mathematics}
\bvolume{58}
\bpages{1--22}.
\end{barticle}
\endbibitem

\bibitem[\protect\citeauthoryear{Liu et~al.}{2020}]{liu2020model}
\begin{barticle}[author]
\bauthor{\bsnm{Liu},~\bfnm{Wanjun}\binits{W.}},
  \bauthor{\bsnm{Ke},~\bfnm{Yuan}\binits{Y.}},
  \bauthor{\bsnm{Liu},~\bfnm{Jingyuan}\binits{J.}} \AND
  \bauthor{\bsnm{Li},~\bfnm{Runze}\binits{R.}}
(\byear{2020}).
\btitle{Model-free feature screening and FDR control with Knockoff features}.
\bjournal{Journal of the American Statistical Association}
\bvolume{0}
\bpages{1--16}.
\end{barticle}
\endbibitem

\bibitem[\protect\citeauthoryear{Lugosi and Vayatis}{2004}]{lugosi2004bayes}
\begin{barticle}[author]
\bauthor{\bsnm{Lugosi},~\bfnm{G{\'a}bor}\binits{G.}} \AND
  \bauthor{\bsnm{Vayatis},~\bfnm{Nicolas}\binits{N.}}
(\byear{2004}).
\btitle{On the Bayes-risk consistency of regularized boosting methods}.
\bjournal{The Annals of statistics}
\bvolume{32}
\bpages{30--55}.
\end{barticle}
\endbibitem

\bibitem[\protect\citeauthoryear{Ma et~al.}{2019}]{ma2019flexible}
\begin{barticle}[author]
\bauthor{\bsnm{Ma},~\bfnm{Jiaqi}\binits{J.}},
  \bauthor{\bsnm{Tang},~\bfnm{Weijing}\binits{W.}},
  \bauthor{\bsnm{Zhu},~\bfnm{Ji}\binits{J.}} \AND
  \bauthor{\bsnm{Mei},~\bfnm{Qiaozhu}\binits{Q.}}
(\byear{2019}).
\btitle{A flexible generative framework for graph-based semi-supervised
  learning}.
\bjournal{Advances in Neural Information Processing Systems}
\bvolume{32}
\bpages{3281--3290}.
\end{barticle}
\endbibitem

\bibitem[\protect\citeauthoryear{Mai and Zou}{2013}]{mai2013kolmogorov}
\begin{barticle}[author]
\bauthor{\bsnm{Mai},~\bfnm{Qing}\binits{Q.}} \AND
  \bauthor{\bsnm{Zou},~\bfnm{Hui}\binits{H.}}
(\byear{2013}).
\btitle{The Kolmogorov filter for variable screening in high-dimensional binary
  classification}.
\bjournal{Biometrika}
\bvolume{100}
\bpages{229--234}.
\end{barticle}
\endbibitem

\bibitem[\protect\citeauthoryear{Mai and Zou}{2015}]{mai2015fused}
\begin{barticle}[author]
\bauthor{\bsnm{Mai},~\bfnm{Qing}\binits{Q.}} \AND
  \bauthor{\bsnm{Zou},~\bfnm{Hui}\binits{H.}}
(\byear{2015}).
\btitle{The fused Kolmogorov filter: A nonparametric model-free screening
  method}.
\bjournal{The Annals of Statistics}
\bvolume{43}
\bpages{1471--1497}.
\end{barticle}
\endbibitem

\bibitem[\protect\citeauthoryear{Maqableh et~al.}{2021}]{maqableh2021effect}
\begin{barticle}[author]
\bauthor{\bsnm{Maqableh},~\bfnm{M}\binits{M.}},
  \bauthor{\bsnm{Abuhashesh},~\bfnm{M}\binits{M.}},
  \bauthor{\bsnm{Dahabiyeh},~\bfnm{L}\binits{L.}},
  \bauthor{\bsnm{Nawayseh},~\bfnm{M}\binits{M.}} \AND
  \bauthor{\bsnm{Masadeh},~\bfnm{R}\binits{R.}}
(\byear{2021}).
\btitle{The effect of Facebook users’ satisfaction and trust on stickiness:
  the role of perceived values}.
\bjournal{International Journal of Data and Network Science}
\bvolume{5}
\bpages{245--256}.
\end{barticle}
\endbibitem

\bibitem[\protect\citeauthoryear{Milli et~al.}{2025}]{milli2025engagement}
\begin{barticle}[author]
\bauthor{\bsnm{Milli},~\bfnm{Smitha}\binits{S.}},
  \bauthor{\bsnm{Carroll},~\bfnm{Micah}\binits{M.}},
  \bauthor{\bsnm{Wang},~\bfnm{Yike}\binits{Y.}},
  \bauthor{\bsnm{Pandey},~\bfnm{Sashrika}\binits{S.}},
  \bauthor{\bsnm{Zhao},~\bfnm{Sebastian}\binits{S.}} \AND
  \bauthor{\bsnm{Dragan},~\bfnm{Anca~D}\binits{A.~D.}}
(\byear{2025}).
\btitle{Engagement, user satisfaction, and the amplification of divisive
  content on social media}.
\bjournal{PNAS nexus}
\bvolume{4}
\bpages{pgaf062}.
\end{barticle}
\endbibitem

\bibitem[\protect\citeauthoryear{Mukherjee et~al.}{2021}]{mukherjee2021high}
\begin{barticle}[author]
\bauthor{\bsnm{Mukherjee},~\bfnm{Somabha}\binits{S.}},
  \bauthor{\bsnm{Niu},~\bfnm{Ziang}\binits{Z.}},
  \bauthor{\bsnm{Halder},~\bfnm{Sagnik}\binits{S.}},
  \bauthor{\bsnm{Bhattacharya},~\bfnm{Bhaswar~B}\binits{B.~B.}} \AND
  \bauthor{\bsnm{Michailidis},~\bfnm{George}\binits{G.}}
(\byear{2021}).
\btitle{High dimensional logistic regression under network dependence}.
\bjournal{arXiv preprint arXiv:2110.03200}.
\end{barticle}
\endbibitem

\bibitem[\protect\citeauthoryear{Pan and Wang}{2015}]{pan2015note}
\begin{barticle}[author]
\bauthor{\bsnm{Pan},~\bfnm{Rui}\binits{R.}} \AND
  \bauthor{\bsnm{Wang},~\bfnm{HanSheng}\binits{H.}}
(\byear{2015}).
\btitle{A note on testing conditional independence for social network
  analysis}.
\bjournal{Science China Mathematics}
\bvolume{58}
\bpages{1179--1190}.
\end{barticle}
\endbibitem

\bibitem[\protect\citeauthoryear{Pan et~al.}{2019a}]{pan2019ball}
\begin{barticle}[author]
\bauthor{\bsnm{Pan},~\bfnm{Wenliang}\binits{W.}},
  \bauthor{\bsnm{Wang},~\bfnm{Xueqin}\binits{X.}},
  \bauthor{\bsnm{Zhang},~\bfnm{Heping}\binits{H.}},
  \bauthor{\bsnm{Zhu},~\bfnm{Hongtu}\binits{H.}} \AND
  \bauthor{\bsnm{Zhu},~\bfnm{Jin}\binits{J.}}
(\byear{2019}a).
\btitle{Ball covariance: A generic measure of dependence in banach space}.
\bjournal{Journal of the American Statistical Association}
\bvolume{529}
\bpages{307--317}.
\end{barticle}
\endbibitem

\bibitem[\protect\citeauthoryear{Pan et~al.}{2019b}]{pan2018generic}
\begin{barticle}[author]
\bauthor{\bsnm{Pan},~\bfnm{Wenliang}\binits{W.}},
  \bauthor{\bsnm{Wang},~\bfnm{Xueqin}\binits{X.}},
  \bauthor{\bsnm{Xiao},~\bfnm{Weinan}\binits{W.}} \AND
  \bauthor{\bsnm{Zhu},~\bfnm{Hongtu}\binits{H.}}
(\byear{2019}b).
\btitle{A generic sure independence screening procedure}.
\bjournal{Journal of the American Statistical Association}
\bvolume{114}
\bpages{928--937}.
\end{barticle}
\endbibitem

\bibitem[\protect\citeauthoryear{Pang et~al.}{2020}]{pang2020improving}
\begin{barticle}[author]
\bauthor{\bsnm{Pang},~\bfnm{Ming}\binits{M.}},
  \bauthor{\bsnm{Ting},~\bfnm{Kai~Ming}\binits{K.~M.}},
  \bauthor{\bsnm{Zhao},~\bfnm{Peng}\binits{P.}} \AND
  \bauthor{\bsnm{Zhou},~\bfnm{Zhi-Hua}\binits{Z.-H.}}
(\byear{2020}).
\btitle{Improving deep forest by screening}.
\bjournal{IEEE Transactions on Knowledge and Data Engineering}
\bvolume{34}
\bpages{4298--4312}.
\end{barticle}
\endbibitem

\bibitem[\protect\citeauthoryear{Sit, Ying and Yu}{2021}]{sit2021event}
\begin{barticle}[author]
\bauthor{\bsnm{Sit},~\bfnm{T}\binits{T.}},
  \bauthor{\bsnm{Ying},~\bfnm{Z}\binits{Z.}} \AND
  \bauthor{\bsnm{Yu},~\bfnm{Yi}\binits{Y.}}
(\byear{2021}).
\btitle{Event history analysis of dynamic networks}.
\bjournal{Biometrika}
\bvolume{108}
\bpages{223--230}.
\end{barticle}
\endbibitem

\bibitem[\protect\citeauthoryear{Su and Wu}{2019}]{su2019learning}
\begin{binproceedings}[author]
\bauthor{\bsnm{Su},~\bfnm{Bing}\binits{B.}} \AND
  \bauthor{\bsnm{Wu},~\bfnm{Ying}\binits{Y.}}
(\byear{2019}).
\btitle{Learning distance for sequences by learning a ground metric}.
In \bbooktitle{International Conference on Machine Learning}
\bpages{6015--6025}.
\bpublisher{PMLR}.
\end{binproceedings}
\endbibitem

\bibitem[\protect\citeauthoryear{Tong et~al.}{2023}]{tong2023model}
\begin{barticle}[author]
\bauthor{\bsnm{Tong},~\bfnm{Zhaoxue}\binits{Z.}},
  \bauthor{\bsnm{Cai},~\bfnm{Zhanrui}\binits{Z.}},
  \bauthor{\bsnm{Yang},~\bfnm{Songshan}\binits{S.}} \AND
  \bauthor{\bsnm{Li},~\bfnm{Runze}\binits{R.}}
(\byear{2023}).
\btitle{Model-free conditional feature screening with FDR control}.
\bjournal{Journal of the American Statistical Association}
\bvolume{118}
\bpages{2575--2587}.
\end{barticle}
\endbibitem

\bibitem[\protect\citeauthoryear{Wang, Aribarg and
  Atchad{\'e}}{2013}]{wang2013modeling}
\begin{barticle}[author]
\bauthor{\bsnm{Wang},~\bfnm{Jing}\binits{J.}},
  \bauthor{\bsnm{Aribarg},~\bfnm{Anocha}\binits{A.}} \AND
  \bauthor{\bsnm{Atchad{\'e}},~\bfnm{Yves~F}\binits{Y.~F.}}
(\byear{2013}).
\btitle{Modeling choice interdependence in a social network}.
\bjournal{Marketing Science}
\bvolume{32}
\bpages{977--997}.
\end{barticle}
\endbibitem

\bibitem[\protect\citeauthoryear{Wasserman et~al.}{1994}]{wasserman1994social}
\begin{bbook}[author]
\bauthor{\bsnm{Wasserman},~\bfnm{Stanley}\binits{S.}},
  \bauthor{\bsnm{Faust},~\bfnm{Katherine}\binits{K.}} \betal{et~al.}
(\byear{1994}).
\btitle{Social network analysis: Methods and applications}.
\bpublisher{Cambridge university press}.
\end{bbook}
\endbibitem

\bibitem[\protect\citeauthoryear{Wu and Leng}{2023}]{wu2023random}
\begin{barticle}[author]
\bauthor{\bsnm{Wu},~\bfnm{Weichi}\binits{W.}} \AND
  \bauthor{\bsnm{Leng},~\bfnm{Chenlei}\binits{C.}}
(\byear{2023}).
\btitle{A Random Graph-based Autoregressive Model for Networked Time Series}.
\bjournal{arXiv preprint arXiv:2309.08488}.
\end{barticle}
\endbibitem

\bibitem[\protect\citeauthoryear{Xie et~al.}{2020}]{xie2020category}
\begin{barticle}[author]
\bauthor{\bsnm{Xie},~\bfnm{Jinhan}\binits{J.}},
  \bauthor{\bsnm{Lin},~\bfnm{Yuanyuan}\binits{Y.}},
  \bauthor{\bsnm{Yan},~\bfnm{Xiaodong}\binits{X.}} \AND
  \bauthor{\bsnm{Tang},~\bfnm{Niansheng}\binits{N.}}
(\byear{2020}).
\btitle{Category-adaptive variable screening for ultra-high dimensional
  heterogeneous categorical data}.
\bjournal{Journal of the American Statistical Association}
\bvolume{115}
\bpages{747--760}.
\end{barticle}
\endbibitem

\bibitem[\protect\citeauthoryear{Yan et~al.}{2019}]{yan2019statistical}
\begin{barticle}[author]
\bauthor{\bsnm{Yan},~\bfnm{Ting}\binits{T.}},
  \bauthor{\bsnm{Jiang},~\bfnm{Binyan}\binits{B.}},
  \bauthor{\bsnm{Fienberg},~\bfnm{Stephen~E}\binits{S.~E.}} \AND
  \bauthor{\bsnm{Leng},~\bfnm{Chenlei}\binits{C.}}
(\byear{2019}).
\btitle{Statistical inference in a directed network model with covariates}.
\bjournal{Journal of the American Statistical Association}
\bvolume{114}
\bpages{857--868}.
\end{barticle}
\endbibitem

\bibitem[\protect\citeauthoryear{Yu et~al.}{2021}]{yu2021collaborative}
\begin{barticle}[author]
\bauthor{\bsnm{Yu},~\bfnm{Xianshi}\binits{X.}},
  \bauthor{\bsnm{Li},~\bfnm{Ting}\binits{T.}},
  \bauthor{\bsnm{Ying},~\bfnm{Ningchen}\binits{N.}} \AND
  \bauthor{\bsnm{Jing},~\bfnm{Bing-Yi}\binits{B.-Y.}}
(\byear{2021}).
\btitle{Collaborative Filtering with Awareness of Social Networks}.
\bjournal{Journal of Business \& Economic Statistics}
\bpages{1--13}.
\end{barticle}
\endbibitem

\bibitem[\protect\citeauthoryear{Zhang, Xu and Zhu}{2022}]{zhang2022joint}
\begin{barticle}[author]
\bauthor{\bsnm{Zhang},~\bfnm{Xuefei}\binits{X.}},
  \bauthor{\bsnm{Xu},~\bfnm{Gongjun}\binits{G.}} \AND
  \bauthor{\bsnm{Zhu},~\bfnm{Ji}\binits{J.}}
(\byear{2022}).
\btitle{Joint latent space models for network data with high-dimensional node
  variables}.
\bjournal{Biometrika}
\bvolume{109}
\bpages{707--720}.
\end{barticle}
\endbibitem

\bibitem[\protect\citeauthoryear{Zhang and Zhu}{2024}]{zhang2024projective}
\begin{barticle}[author]
\bauthor{\bsnm{Zhang},~\bfnm{Yaowu}\binits{Y.}} \AND
  \bauthor{\bsnm{Zhu},~\bfnm{Liping}\binits{L.}}
(\byear{2024}).
\btitle{Projective independence tests in high dimensions: the curses and the
  cures}.
\bjournal{Biometrika}
\bvolume{111}
\bpages{1013--1027}.
\end{barticle}
\endbibitem

\bibitem[\protect\citeauthoryear{Zhang et~al.}{2017}]{zhang2017influence}
\begin{barticle}[author]
\bauthor{\bsnm{Zhang},~\bfnm{Mingli}\binits{M.}},
  \bauthor{\bsnm{Guo},~\bfnm{Lingyun}\binits{L.}},
  \bauthor{\bsnm{Hu},~\bfnm{Mu}\binits{M.}} \AND
  \bauthor{\bsnm{Liu},~\bfnm{Wenhua}\binits{W.}}
(\byear{2017}).
\btitle{Influence of customer engagement with company social networks on
  stickiness: Mediating effect of customer value creation}.
\bjournal{International Journal of Information Management}
\bvolume{37}
\bpages{229--240}.
\end{barticle}
\endbibitem

\bibitem[\protect\citeauthoryear{Zhang et~al.}{2020}]{zhang2020logistic}
\begin{barticle}[author]
\bauthor{\bsnm{Zhang},~\bfnm{Xu}\binits{X.}},
  \bauthor{\bsnm{Pan},~\bfnm{Rui}\binits{R.}},
  \bauthor{\bsnm{Guan},~\bfnm{Guoyu}\binits{G.}},
  \bauthor{\bsnm{Zhu},~\bfnm{Xuening}\binits{X.}} \AND
  \bauthor{\bsnm{Wang},~\bfnm{Hansheng}\binits{H.}}
(\byear{2020}).
\btitle{Logistic regression with network structure}.
\bjournal{Statistica Sinica}
\bvolume{30}
\bpages{673--693}.
\end{barticle}
\endbibitem

\bibitem[\protect\citeauthoryear{Zhang et~al.}{2021}]{zhang2021depth}
\begin{barticle}[author]
\bauthor{\bsnm{Zhang},~\bfnm{Xu}\binits{X.}},
  \bauthor{\bsnm{Tian},~\bfnm{Yahui}\binits{Y.}},
  \bauthor{\bsnm{Guan},~\bfnm{Guoyu}\binits{G.}} \AND
  \bauthor{\bsnm{Gel},~\bfnm{Yulia~R}\binits{Y.~R.}}
(\byear{2021}).
\btitle{Depth-based classification for relational data with multiple
  attributes}.
\bjournal{Journal of Multivariate Analysis}
\bvolume{184}
\bpages{104732}.
\end{barticle}
\endbibitem

\bibitem[\protect\citeauthoryear{Zhong et~al.}{2023}]{zhong2023feature}
\begin{barticle}[author]
\bauthor{\bsnm{Zhong},~\bfnm{Wei}\binits{W.}},
  \bauthor{\bsnm{Qian},~\bfnm{Chen}\binits{C.}},
  \bauthor{\bsnm{Liu},~\bfnm{Wanjun}\binits{W.}},
  \bauthor{\bsnm{Zhu},~\bfnm{Liping}\binits{L.}} \AND
  \bauthor{\bsnm{Li},~\bfnm{Runze}\binits{R.}}
(\byear{2023}).
\btitle{Feature screening for interval-valued response with application to
  study association between posted salary and required skills}.
\bjournal{Journal of the American Statistical Association}
\bvolume{118}
\bpages{805--817}.
\end{barticle}
\endbibitem

\bibitem[\protect\citeauthoryear{Zhu et~al.}{2017}]{zhu2017network}
\begin{barticle}[author]
\bauthor{\bsnm{Zhu},~\bfnm{Xuening}\binits{X.}},
  \bauthor{\bsnm{Pan},~\bfnm{Rui}\binits{R.}},
  \bauthor{\bsnm{Li},~\bfnm{Guodong}\binits{G.}},
  \bauthor{\bsnm{Liu},~\bfnm{Yuewen}\binits{Y.}},
  \bauthor{\bsnm{Wang},~\bfnm{Hansheng}\binits{H.}} \betal{et~al.}
(\byear{2017}).
\btitle{Network vector autoregression}.
\bjournal{The Annals of Statistics}
\bvolume{45}
\bpages{1096--1123}.
\end{barticle}
\endbibitem

\end{thebibliography}

\end{document}